%% file: main.tex
\documentclass{article}

\usepackage{microtype}
\usepackage{graphicx}
\usepackage{subfigure}
\usepackage{booktabs} % for professional tables

\usepackage{xurl}
\usepackage{etoolbox}

\usepackage{tikz}
\usepackage{pgfplots}
\usepackage{pgfplotstable}
\usepackage{etoolbox}

\usepackage{xcolor}
\usepackage{todonotes}

\usetikzlibrary{matrix,calc,backgrounds,fit,arrows.meta,shapes.geometric,patterns.meta}
\usepgfplotslibrary{groupplots}

\usepackage{multirow}

\usepackage{hyperref}

\usepackage{xspace}

\usepackage[accepted]{mlsys2025}

\makeatletter
\renewcommand{\Notice@String}{}
\makeatother

\mlsystitlerunning{\name: Directly Invoking Models Without Intensive Tuning}
\input{macros}

\begin{document}

\twocolumn[
\mlsystitle{\name: Directly Invoking Models Without Intensive Tuning}
% \mlsystitle{Why learn when you can guess?} %General-purpose Uncertainty in Engineering Systems Selection}

\begin{mlsysauthorlist}
\mlsysauthor{Saarth Deshpande}{ut}
\mlsysauthor{Neeraja J. Yadwadkar}{ut}
\mlsysauthor{Venkat Arun}{ut}
\end{mlsysauthorlist}

\mlsyscorrespondingauthor{Saarth Deshpande}{saarth@utexas.edu}

\mlsysaffiliation{ut}{The University of Texas at Austin}

\mlsyskeywords{Resource Management, Large Language Models, Decision Models, Systems Optimization, Autoscaling, DRAM Sizing, Video Compression}

\vskip 0.3in

\input{abstract}
]

\printAffiliationsAndNotice{}

\input{intro}

\input{approach}

\input{impl}

\input{eval}

\input{discussion}

\input{related-work}

\input{conclusion}


\bibliography{refs}
\bibliographystyle{mlsys2025}

\section*{AI Usage Disclosure}
Generative AI tools, specifically ChatGPT and Claude Code, were used to assist with literature searches and citation verification, code development and debugging, and drafting and editing portions of the manuscript. The models evaluated in this paper were also used to make resource-management decisions in the experiments, as described in the paper. The key ideas and experimental design were developed by the human authors, and experimental results were obtained from actual system executions. The authors reviewed and edited all AI-assisted outputs and take full responsibility for the paper's content, including the accuracy of its claims and citations.

% \appendix
% \section{}

\end{document}

%% file: macros.tex
\newcommand{\name}{\textsc{Dimwit}\xspace}

\newif\ifshowcomments
\showcommentsfalse

%% file: abstract.tex
\begin{abstract}
    Resource managers typically encode what to do in a task-specific policy---a heuristic, an optimization procedure, or a model trained for the task---that must be developed before use and may need revision as the workload or system changes. We propose bypassing policy development and instead asking pretrained models to make resource-management decisions directly, given the current system state, desired outcome, and available choices. We evaluate this approach on microservice autoscaling, DRAM sizing, and video encoding, using a frontier LLM and two small decision models. On two microservice applications, a small decision model meets latency targets while using within 10\% of the replicas allocated by policy-based autoscalers and within 13\% of the frontier LLM. In DRAM sizing, a small decision model reduces DRAM usage by up to 27.5\% relative to static provisioning while maintaining at least 95\% of baseline performance. In video encoding, it completes encodes up to 4.3$\times$ faster than the frontier LLM, with an imperceptible 1.1-point decrease in Video Multi-Method Assessment Fusion (VMAF), a metric of perceptual video quality. Making individual resource-management decisions may be considerably simpler than developing a general strategy for making them.
\end{abstract}

%% file: intro.tex
\section{Introduction}

Many systems problems invite an unreasonable thought: “If I could just sit in the control loop, I would know what to do.” Resource management requires meeting performance targets without wasting resources, even as workloads and system conditions change. Say a service is overloaded while its neighbors are idle: give that service more replicas. Say an application gains little from additional fast memory: try giving it less. The intuition may be imperfect, but it is often easy to express. We might even want to steer a running system by saying “this is the service that’s misbehaving” or “err on the side of lower local DRAM usage,” and have it adjust accordingly. The obvious problem is that nobody wants to spend their day being a resource allocator.

These decisions are typically automated using a \emph{resource-management policy}: a mapping from system measurements to actions. A hand-tuned threshold is a policy; so is a decision tree, an optimization procedure, or a neural network trained with reinforcement learning. They differ in how the mapping is obtained, but each must be designed, tuned, or trained for the task before the system can act, and may require revision as the workload or system changes. A simple policy, such as scaling replicas when CPU utilization exceeds a fixed threshold, may waste resources or fail to meet performance targets~\cite{deathstarbench}. Making these policies effective often requires domain expertise, application-specific profiling, and repeated experimentation to balance resource consumption against performance targets. More recently, LLMs have been used to generate and refine such heuristics~\cite{funsearch,vulcan}, but the output is still a policy that must be produced and validated before use. This paper asks whether we can skip developing a task-specific policy altogether.

For some applications, we can instead ask a language model to make each decision directly. This is practical when the system can tolerate the LLM's inference latency and the inference cost is offset by the number of resources being managed. Decision models such as Jev~\cite{jev} and Laya~\cite{laya} make this practical for many more applications: a decision takes 0.1--0.2\,s with Jev or with Laya on a GPU (2--3\,s for Laya on a CPU with 16 threads), against 3--4\,s for Opus 5.5 at low reasoning effort and 7--8\,s at high. The proposition is slightly absurd, but concrete: instead of writing a policy that captures what we would do, describe the situation and ask what to do next. That is, make the relevant information explicit without prescribing how the model should use it to make a decision.

We investigate whether pretrained models---general-purpose LLMs and small decision models---can make resource-management decisions that minimize resource use while meeting performance targets, without developing a policy for each task. We also ask how sophisticated a model needs to be to make these decisions.
We evaluate three models: Claude Opus 5.5~\cite{opus-5-5}, a general-purpose LLM, Jev~\cite{jev}, a proprietary decision model, and Laya~\cite{laya}, an open-weight 421M-parameter decision model. We refer to resource managers built using these models as \name resource managers.

Our evaluation covers three representative use cases. First, we consider microservice applications, where we perform \emph{horizontal autoscaling}: adjusting the number of replicas of each service to reduce resource use while meeting a latency target. Second, we consider applications that use both fast local DRAM and remote Compute Express Link (CXL) memory. Here, we dynamically adjust the amount of DRAM allocated to an application while keeping its slowdown within a limit. Third, we consider \emph{video compression}, where we adjust compression settings to reduce encoding time while meeting a visual-quality target.

As we show in \S\ref{sec:eval}, even the small decision models perform competitively with Claude Opus 5.5 and specialized resource-management systems across all three tasks, without task-specific training or profiling. In DRAM sizing and video encoding, Laya makes more aggressive choices than Opus 5.5, allocating up to approximately 19\% less DRAM while meeting performance targets on six of eight workloads, and reducing encoding time by up to 4.3$\times$ within a 1.1-point reduction in perceptual video quality on a 100-point scale, which is an imperceptible difference~\cite{vmaf}. 

Our studies yield several conclusions. First, these problems are simple enough that large models do not always offer a meaningful advantage. A 421M-parameter open-source model performs competitively with a frontier LLM.  Second, while these models do not dramatically outperform hand-designed heuristics, and sometimes perform worse, they achieve comparable results without requiring the development of task-specific policies. They allow the developer to ``talk'' to their systems to tweak their behavior. This makes them a practical alternative when developers need a quick heuristic but lack the time to translate their intuition into a control algorithm, provided the application can tolerate the inference overhead. Having a model-in-the-loop ensures common sense and prevents catastrophically bad performance.

%% file: approach.tex
\section{Is the decision the easy part?}

Managing resources like CPU and memory typically requires task-specific mechanisms to decide what actions to take, whether through learned models, optimization algorithms, or hand-tuned heuristics. In each case, the intuition about what to do must first be made precise---written as a rule, posed as an optimization, or learned from data---before the system can act on it. We explore an alternative: describe the current system state, the desired outcome, and the available choices, and let a language model decide how to allocate resources. For this to work, the model must have enough relevant information to evaluate the tradeoffs between available choices. The challenge thus shifts from specifying how to make decisions to specifying the information needed to make them.

Our main design question was simply: “If I could only see what the model sees, would I have enough information to make a decision?” When the answer was no, we added the missing information. Why are humans, LLMs and decision models comparable? Because many of the problems are simple enough that intelligence does not offer a meaningful advantage. Cleverness is necessary when designing an algorithm that must handle every eventuality. That is why many of these have been rich areas of research. Cleverness is less important when making the decision directly in the loop. This is the key insight of the paper.

When the model did not make the desired trade-off, we could adjust the model’s behavior with instructions such as “err on the side of lower local memory usage,” without rewriting a heuristic or retraining a policy. 

All three tasks pose the model the same problem: choose among discrete configurations to minimize a cost while meeting a performance target. In horizontal autoscaling for microservices, this means minimizing the number of replicas while meeting latency targets; in DRAM sizing, minimizing allocated DRAM while maintaining performance; and in video encoding, minimizing encoding time while preserving video quality. Only the system state, available configurations, objectives, and constraints differ.

Yet sharing the same structure does not make these decisions straightforward. For example, the effect of adding replicas depends on workload and interactions among services, while that of reducing DRAM depends on memory access patterns. These relationships are precisely what task-specific mechanisms attempt to capture. How, then, can a language model make useful decisions without first learning or profiling them?

We conjecture that useful resource-management decisions can be made from the system's observed behavior, without learning either a model of the system or a task-specific decision policy. The current resource allocation, recent changes in workload and performance, and distance from the target may already be enough to determine what to adjust. What is difficult to encode as a general decision-making strategy may be considerably simpler to decide one instance at a time. If individual decisions are indeed simpler than learning a general decision-making strategy, how sophisticated does the model making them need to be?

We investigate this using a frontier LLM~\cite{opus-5-5} and two small decision models~\cite{laya, jev}, asking each to solve the same resource-management problems. The next section describes what information each model receives, what choices it makes, and how its decisions are applied.

%% file: impl.tex
\section{Implementation}

\input{impl_autoscaling}

\input{tables/ffmpeg-params}

\input{impl_dram}

\input{impl_ffmpeg}

%% file: impl_autoscaling.tex
\subsection{Horizontal Autoscaling}

Microservice applications consist of multiple interacting services whose resource demands change with workload~\cite{deathstarbench}. Horizontal autoscaling adjusts the number of replicas of each service in response to changing workloads while satisfying application latency Service-Level Objectives (SLOs)~\cite{erlang}. However, determining which services to scale and by how much requires reasoning about how different resource allocations affect end-to-end performance~\cite{sinan-delimitrou,erlang}. 

We integrate the \name autoscalers with Kubernetes by replacing the scaling policy of its Horizontal Pod Autoscaler (HPA)~\cite{k8s-hpa}, the most widely used autoscaler today~\cite{datadog-containers-serverless-2025}. Every 15 seconds, the pretrained models determine the replica counts for the application's microservices. We apply these decisions by setting the minimum and maximum replica counts of the corresponding HPA objects to the selected values, allowing Kubernetes to carry out the scaling operations (see Figure~\ref{fig:k8s-stack}).

When autoscaling, each \name model is provided the following state: current replica count of a service, its CPU and memory utilization, time since the last replica change, and its average replica startup time. We also provide the latency SLO, per-endpoint request rates and p95 latencies, and the change in each endpoint's request rate over the past 60 seconds. In addition to these inputs, Opus 5.5 receives a history of its previous scaling decisions and the reasoning it provided for those decisions, including its assessment of workload trends and resource requirements.

To manage the overhead of model inference, we batch the states of multiple services into a single request to Laya, since invoking the self-hosted model separately for each service would be computationally expensive. Unlike Opus 5.5, which jointly determines the replica counts of all services in a single decision, Laya's batching groups independent per-service decisions into one API call. Jev does not support batched requests, so we issue one API call per service every 15\,s.

We ask Laya and Jev to choose among seven scaling factors: 0.25, 0.5, 0.75, 1, 1.5, 2, and 2.5 times the current replica count, subject to a minimum of 1 and a maximum of 50 replicas per service. Opus 5.5 instead selects the absolute replica count of every service in a single decision. We instruct all three models to select the minimum number of replicas needed to meet the p95 latency SLO under the load expected during replica startup. We also instruct them to scale out proactively when increased load is anticipated, while otherwise preferring to keep or reduce replicas instead of overprovisioning.

\input{figures/impl_autoscaler}

%% file: figures/impl_autoscaler.tex
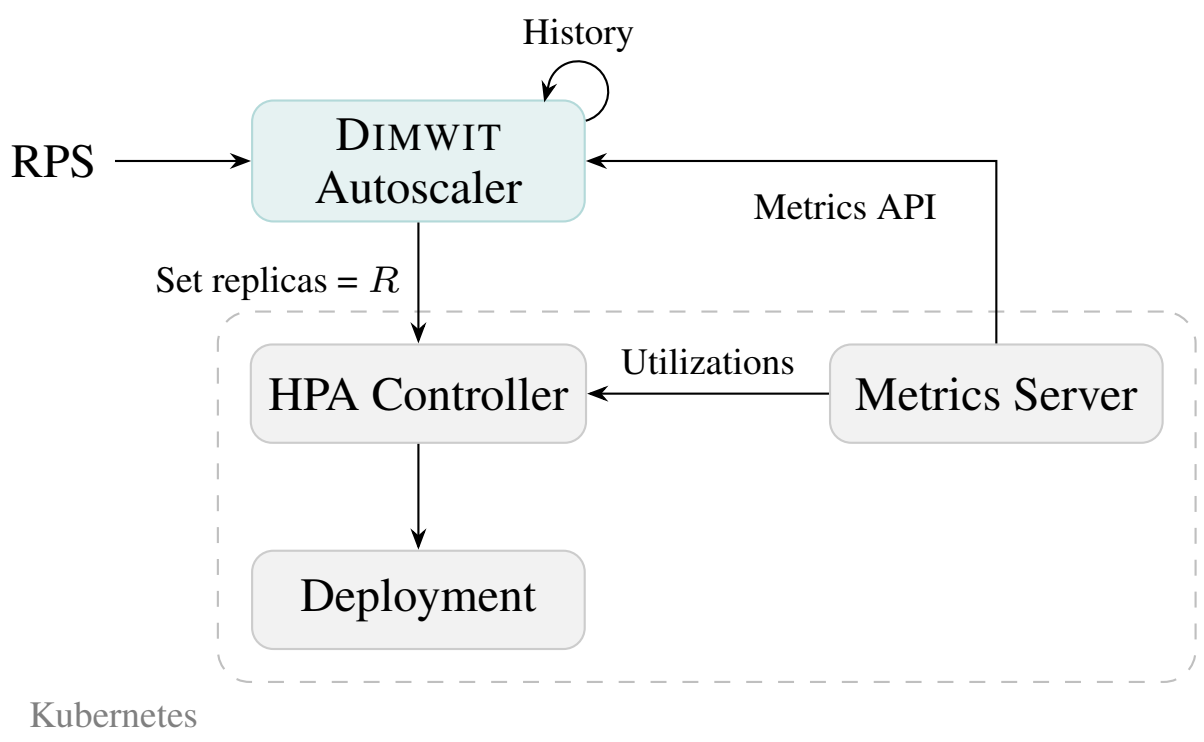
\begin{figure}[t]
\centering
\begin{tikzpicture}[
    box/.style={rectangle, rounded corners=4pt, draw, minimum width=2.2cm,
                minimum height=0.65cm, align=center, font=\small},
    graybox/.style={box, fill=gray!10, draw=gray!40},
    tealbox/.style={box, fill=teal!10, draw=teal!30},
    dashedbox/.style={rectangle, rounded corners=6pt, draw=gray!50, dashed,
                      inner sep=6pt},
    arr/.style={-{Stealth[length=4pt]}, thin},
    lbl/.style={font=\scriptsize},
]

\node[tealbox, minimum width=2.2cm] (wi) {\name\\Autoscaler};
\node[lbl, left=0.9cm of wi] (rps) {\small RPS};
\node[graybox, below=0.8cm of wi] (hpa) {HPA Controller};
\node[graybox, right=1.6cm of hpa] (metrics) {Metrics Server};
\node[graybox, below=0.7cm of hpa] (deploy) {Deployment};

\begin{scope}[on background layer]
  \node[dashedbox, fit=(hpa)(metrics)(deploy),
        label={[font=\scriptsize\color{gray}]south west:Kubernetes}] (k8s) {};
\end{scope}

%% RPS -> WhatIf
\draw[arr] (rps.east) -- (wi.west);

%% Autoscaler history
\draw[arr] ([xshift=0pt,yshift=-4pt]wi.north east)
    arc[start angle=285, end angle=520, radius=0.2cm]
    -- ++(-1pt,-4pt)
    node[pos=-0.5, above, xshift=5pt, lbl] {History};

%% WhatIf patches HPA
\draw[arr] (wi.south) -- (hpa.north)
    node[midway, left, lbl] {Set replicas = $R$};

%% Metrics → HPA
\draw[arr] (metrics.west) -- (hpa.east)
    node[midway, above, lbl] {Utilizations};

%% Metrics → WhatIf (routed via north to avoid label angle issue)
\draw[arr] (metrics.north) -- ++(0,0.9) |- (wi.east)
    node[pos=0.35, below, xshift=-1.0cm, lbl] {Metrics API};

%% HPA → Deployment
\draw[arr] (hpa.south) -- (deploy.north);

\end{tikzpicture}
\caption{%
   \textbf{Architecture common to the \name autoscalers.} At each tick, the autoscaler reads per-service utilization via the Kubernetes Metrics API and patches the service's HPA object with the desired replica count $R$, causing HPA to adjust the service's Deployment.
}
\label{fig:k8s-stack}
\end{figure}

%% file: tables/ffmpeg-params.tex
\begin{table*}[!t]
\centering
\small
\begin{tabular}{@{}llcc@{}}
\toprule
\textbf{Parameter} & \textbf{Description} & \textbf{Choices} & \textbf{FFmpeg Default} \\
\midrule
\texttt{subme} & Precision of motion estimation and optimization of block encoding & 2, 4, 7, 9 & 7 \\
\texttt{me} & Motion search algorithm & dia, hex, umh & hex \\
\texttt{ref} & Number of reference frames used for prediction & 1, 3, 5 & 3 \\
\texttt{trellis} & Optimization of quantized values & 0, 1, 2 & 1 \\
\texttt{bframes} & Maximum number of consecutive B-frames & 0, 3, 5 & 3 \\
\texttt{b\_strategy} & Algorithm for selecting B-frame positions & 0, 1, 2 & 1 \\
\bottomrule
\end{tabular}
\caption{x264 encoding parameters available to the LLMs, their allowed values, and the defaults under the \texttt{medium} preset.}
\label{tab:encoding-parameters}
\end{table*}

%% file: impl_dram.tex
\subsection{Dynamic DRAM Sizing}

Tiered memory systems combine fast DRAM with slower, cheaper memory such as CXL to reduce memory costs~\cite{tpp2023,cxlpond}. DRAM sizing determines how much memory to allocate to the fast tier while maintaining application performance. 

We implement the \name sizers on top of ARMS~\cite{arms}, a state-of-the-art tiering system. ARMS uses PEBS~\cite{pebs2018} to sample per-page accesses and compute page hotness scores. Periodically, it ranks pages by hotness, migrating the hottest pages to DRAM and the coldest to CXL. Every 30 seconds, each \name sizer makes a decision using the state, sizing options, and prompt described below.

We provide each sizer with the current DRAM allocation, the estimated slowdown at that allocation, and three one-second samples of memory-access behavior. We estimate this slowdown as $(1-s_0)/(1-s)-1$, where $s$ is the stall fraction measured at the current allocation over the latest 30-second interval and $s_0$ is a reference stall fraction measured after warm-up. Each sample includes access rates and estimated latencies for DRAM and CXL, memory-level parallelism, and the estimated fraction of all accesses that would shift from DRAM to CXL if the DRAM allocation were reduced by 4\,GiB. We estimate DRAM and CXL latencies from each tier's measured bandwidth using per-tier bandwidth--latency curves, with linear interpolation within the profiled range.

We ask each sizer to choose among reducing DRAM by 4\,GiB, keeping the current allocation, or increasing DRAM by 4\,GiB, with the goal of minimizing DRAM usage while satisfying the slowdown budget. We conservatively target a slowdown 4 percentage points below the specified budget (e.g., 1\% for a 5\% budget), reserving 2 percentage points for estimation errors and another 2 as additional safety margin. Despite this conservatism, Laya and Jev perform well within the 95\% performance target and Laya saves more DRAM than TierSizer (see~\S\ref{subsec:dram-eval}). Unlike Laya and Jev, Opus 5.5 also receives a history of its previous sizing decisions and reasoning, since its larger context window can accommodate this additional information.

We instruct the \name model to minimize DRAM allocation while meeting the performance target, encouraging it to err on the side of trying a smaller allocation when it expects the target to remain satisfied. Every 30 seconds, it chooses whether to decrease DRAM by 4\,GiB, maintain the current allocation, or increase DRAM by 4\,GiB. The selected allocation is applied through ARMS.

%% file: impl_ffmpeg.tex
\subsection{Video Compression}

\input{figures/ffmpeg-vmaf-range}

At a fixed bitrate, video encoders trade encoding time for video quality~\cite{amirpour2023preset, msu2008x264}. A video is a sequence of frames, each divided into small blocks of pixels. Rather than store each block's pixels directly, the encoder predicts them from neighboring blocks or other frames and stores only the difference. 
For each block, it evaluates candidate predictions and selects the one with the lowest estimated rate-distortion cost~\cite{param-optimization-video-compression}. Evaluating more candidates can improve video quality at a given bitrate but takes longer.

Video quality is commonly measured using Video Multi-Method Assessment Fusion (VMAF)~\cite{vmaf}, a perceptual metric that estimates visual quality on a scale from 0 to 100, with higher scores indicating better quality. As Figure~\ref{fig:encoding-tradeoff} illustrates, encoder configurations span a range of encoding times and video quality at a fixed bitrate. Selecting effective configurations is challenging because encoding parameters affect one another, making it difficult to determine their combined impact on encoding time and video quality~\cite{amirpour2023preset}. Finding effective configurations can require repeated execution: Netflix’s Dynamic Optimizer~\cite{netflixDynamicOptimizer}, for example, encodes each video shot using multiple configurations to identify favorable quality–bitrate tradeoffs.

We investigate whether pretrained models can select encoder configurations without evaluating multiple configurations through repeated encoding. We divide each video into 5-second segments and ask a \name model to select encoding parameters for each segment, with the goal of minimizing encoding time while maintaining an average VMAF across the video at least as high as that achieved by x264’s default configuration.

After encoding each segment, we measure its VMAF and encoding time, which the model uses to select parameters for the next segment. The six parameters and the values the model can select are listed in Table~\ref{tab:encoding-parameters}. 

Unlike autoscaling with the small decision models and DRAM sizing, which select changes relative to the current configuration (scaling factors and $\pm$4\,GiB, respectively), video encoding selects absolute values for each encoder parameter. This lets the model choose any allowed value, regardless of the current setting. 

The model receives the target VMAF, the average VMAF achieved so far, the minimum average required over the remaining segments to meet the target, the remaining video duration, the current settings, and the settings, VMAF, and CPU time of the last five segments. We also provide a short description of each setting’s effect on encoding speed and quality.

%% file: figures/ffmpeg-vmaf-range.tex
\begin{figure*}[!t]
    \centering
    \includegraphics[width=\textwidth]{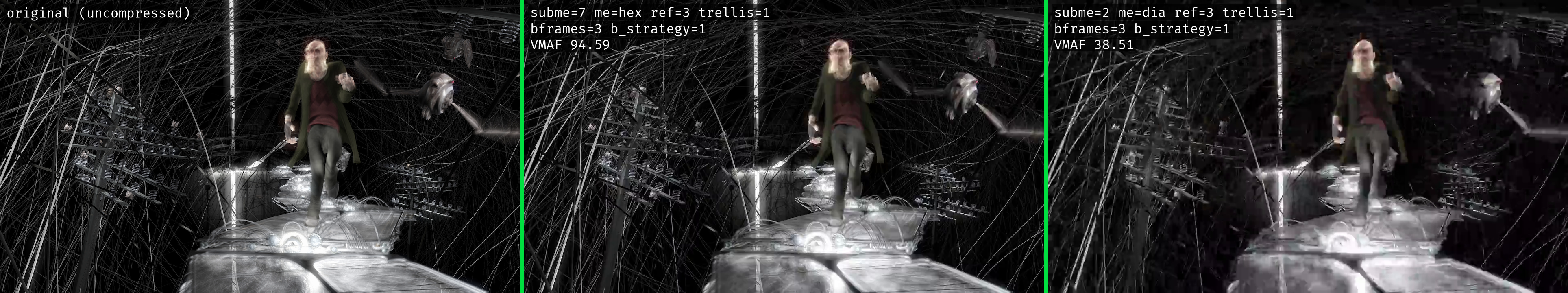}
    \caption{
        Example frame from \emph{Elephants Dream}~\cite{elephants-dream}, shown uncompressed (left) and encoded at the same bitrate using two different \texttt{x264} encoder configurations (center and right). The encoded versions achieve VMAF scores of 94.59 and 38.51, respectively, illustrating how encoder settings affect video quality.
    }
    \label{fig:encoding-tradeoff}
\end{figure*}

%% file: eval.tex
\section{Evaluation}\label{sec:eval}

\input{tables/llm-history}

\input{eval_modelconfig}

\input{eval_autoscaling}

\input{eval_dram}

\input{eval_ffmpeg}

%% file: tables/llm-history.tex
\begin{table}[!t]
\centering
\small
\begin{tabular}{@{}lccc@{}}
\toprule
\textbf{Use case} & \textbf{Laya} & \textbf{Jev} & \textbf{Opus 5.5} \\
\midrule
Autoscaling (past decisions) & -- & -- & 10\\
DRAM Sizing (1\,s access samples) & 3 & 3 & 10\\
Video Compression (past segments) & 2 & 5 & 5\\
\bottomrule
\end{tabular}
\caption{Past observations in each model's state. Opus 5.5 also sees its own past reasoning for those decisions.}
\label{tab:model-history}
\end{table}

%% file: eval_modelconfig.tex
\subsection{Model Configuration.}
We use Laya's \texttt{english} model and \texttt{jev-latest} for Jev, with default input token limits of 512~\cite{laya-huggingface} and 32,000~\cite{jev-models}, respectively. Laya's limit applies to each question--state pair, with up to 192 tokens reserved for the question and its options, which restricts the history it can be given compared with Jev and Opus 5.5. We therefore give Laya a shorter window of recent history than Jev and Opus 5.5 in each of the following use cases. Table~\ref{tab:model-history} summarizes the history provided to each model.

%% file: eval_autoscaling.tex
\subsection{Horizontal Autoscaling}

We evaluate how well \name autoscalers minimize resource allocation while satisfying per-endpoint p95 latency Service-Level Objectives (SLOs) under changing workloads.

\input{eval_autoscaling_setup}

\input{figures/autoscaler-eval}

\input{eval_autoscaling_results}

%% file: eval_autoscaling_setup.tex
\subsubsection{Experimental Setup}

\paragraph{Cluster.}
We run experiments on a CloudLab~\cite{cloudlab} cluster, comprising 15 \texttt{r320} nodes: one control-plane instance, 13 worker instances running Kubernetes deployed with \texttt{kubeadm}, and one dedicated node hosting Laya. Application pods are scheduled only on the worker instances; autoscalers run on the control-plane instance. Opus 5.5 and Jev are accessed through their respective APIs.

\paragraph{Benchmarks.}
We evaluate three microservice applications: TeaStore~\cite{teastore} (7 services), Online Boutique~\cite{online-boutique-github} (11 services), and Social Network~\cite{deathstarbench} (20 services). For Social Network, we exclude MongoDB instances from autoscaling because their Kubernetes \texttt{Deployment} configuration does not support sharding, causing replicas to contend for the same persistent volume. For Online Boutique, we use an external Locust~\cite{locust} instance instead of its bundled load generator.

Kubernetes specifies guaranteed resource allocations through requests and maximum allocations through limits. For Online Boutique, each replica has CPU requests/limits of 1000m/2000m and memory requests/limits of 64/128 MiB, except \texttt{redis-cart}, which retains its defaults of 70m/125m CPU and 200/256 MiB memory. TeaStore uses the same CPU configuration as Online Boutique, with 2 GiB for both memory requests and limits. Social Network uses 500m/1000m CPU and 128/256 MiB memory.

\paragraph{Workloads.}
For each application, we derive replay workloads from Alibaba production traces~\cite{alibaba-2021}, and use the HTTP-entrypoint microservice call rate (MCR) as the workload signal. These HTTP entrypoints provide the most direct measure of offered load. Using call-rate measurements from internal service interactions would instead make the workload signal depend on application-specific implementation details rather than externally imposed demand. Since the original request-rate data is available at 60\,s granularity, we linearly interpolate it to 15\,s resolution for replay. We assign one such workload trace to each application endpoint and rescale the resulting demand to match the capacity of the target deployment. Locust is configured with 100 users and a ramp-up of 1 user/s for TeaStore and Online Boutique, and with 1000 users and a ramp-up of 100 users/s for Social Network.

\paragraph{Baselines.} We compare the \name autoscalers against Erlang~\cite{erlang}, a bandit-learning horizontal autoscaler, Agar~\cite{agar}, which uses analytical models to allocate the minimum number of replicas needed to meet a target SLO, and the Kubernetes Horizontal Pod Autoscaler (HPA)~\cite{k8s-hpa}, configured with hand-tuned utilization thresholds and scaling policies. Given Jev’s low inference latency, we evaluate two variants: Jev-Single, which makes scaling decisions for each microservice using only its local state, and Jev-Batched, which makes a separate scaling decision for each microservice using the shared state of all microservices. Since most prior work focuses on \emph{vertical} autoscaling~\cite{sinan-delimitrou,percs,autothrottle}, we use Erlang and Agar as our research baselines. We tune HPA through extensive experimentation, iteratively adjusting its scaling thresholds based on observed scaling behavior under load, as is common in production autoscalers~\cite{knative-kpa,gcp-autoscaler,aws-autoscaling,gke-metric-autoscaling}. We train both Erlang and Agar’s models on the first six hours of Alibaba traces and evaluate all autoscalers on the subsequent two hours.

%% file: figures/autoscaler-eval.tex
\definecolor{hpaColor}{HTML}{3B82F6}
\definecolor{erlangColor}{HTML}{F97316}
\definecolor{layaColor}{HTML}{16A34A}
\definecolor{opusColor}{HTML}{A855F7}
\definecolor{jevColor}{HTML}{67E8F9}
\definecolor{agarColor}{HTML}{0F766E}

\pgfplotstableread[
    col sep=comma,
    header=true
]{data/autoscaler-eval.csv}\latencydata

% Filled plus marker for Laya.
\pgfdeclareplotmark{filled plus}{%
    \pgfpathrectangle
        {\pgfpoint{-\pgfplotmarksize}{-.333333\pgfplotmarksize}}
        {\pgfpoint{2\pgfplotmarksize}{.666667\pgfplotmarksize}}%
    \pgfpathrectangle
        {\pgfpoint{-.333333\pgfplotmarksize}{-\pgfplotmarksize}}
        {\pgfpoint{.666667\pgfplotmarksize}{2\pgfplotmarksize}}%
    \pgfusepathqfill
}

% Floating-point arithmetic.
\newcommand{\SeriesMath}[2]{%
    \begingroup
    \pgfkeys{/pgf/fpu=true,/pgf/fpu/output format=fixed}%
    \pgfmathparse{#2}%
    \xdef#1{\pgfmathresult}%
    \endgroup
}

% #1 application; #2 autoscaler; #3 color; #4 marker.
\newcommand{\PlotSeriesWithBox}[4]{%
    \gdef\SeriesCoords{}%
    \gdef\SeriesFound{0}%

    \pgfplotstableforeachcolumnelement{app}\of\latencydata
        \as\SeriesApp{%
        \pgfplotstablegetelem{\pgfplotstablerow}{autoscaler}\of\latencydata
        \edef\SeriesAuto{\pgfplotsretval}%

        \ifdefstring{\SeriesApp}{#1}{%
            \ifdefstring{\SeriesAuto}{#2}{%
                \pgfplotstablegetelem
                {\pgfplotstablerow}{replicas}\of\latencydata
            \edef\SeriesX{\pgfplotsretval}%
            
            \pgfplotstablegetelem
                {\pgfplotstablerow}{median_p95_ms}\of\latencydata
                \edef\SeriesY{\pgfplotsretval}%

                \xdef\SeriesCoords{%
                    \SeriesCoords(\SeriesX,\SeriesY)}%

                \ifnum\SeriesFound=0
                    \xdef\SeriesXMin{\SeriesX}%
                    \xdef\SeriesXMax{\SeriesX}%
                    \xdef\SeriesYMin{\SeriesY}%
                    \xdef\SeriesYMax{\SeriesY}%
                    \gdef\SeriesFound{1}%
                \else
                    \SeriesMath{\SeriesXMin}
                        {min(\SeriesXMin,\SeriesX)}%
                    \SeriesMath{\SeriesXMax}
                        {max(\SeriesXMax,\SeriesX)}%
                    \SeriesMath{\SeriesYMin}
                        {min(\SeriesYMin,\SeriesY)}%
                    \SeriesMath{\SeriesYMax}
                        {max(\SeriesYMax,\SeriesY)}%
                \fi
            }{}%
        }{}%
    }%

    \ifnum\SeriesFound=1
        \SeriesMath{\BoxLeft}{max(0,\SeriesXMin-1)}%
        \SeriesMath{\BoxRight}{min(420,\SeriesXMax+1)}%
        \SeriesMath{\BoxBottom}{max(0,\SeriesYMin-10)}%
        \SeriesMath{\BoxTop}{min(70000,\SeriesYMax+10)}%
        
        \def\SeriesMarkSize{2.7pt}%

        \ifstrequal{#2}{Laya}{%
            \def\SeriesMarkSize{3.2pt}%
        }{}%

        \def\SeriesFill{#3}%
        \ifstrequal{#2}{jevSingle}{%
            \def\SeriesFill{white}%
        }{}%
        \ifstrequal{#2}{opusLow}{%
            \def\SeriesFill{white}%
        }{}%

        % Freeze coordinates before PGFPlots defers rendering.
        % Each series keeps its own box; points are never shifted.
        \edef\RenderSeries{%
            \noexpand\addplot[
                no marks, draw=#3, fill=none,
                line width=0.8pt,
                dash pattern=on 4pt off 3.2pt,
                on layer=axis background, forget plot
            ] coordinates {
                (\BoxLeft,\BoxBottom)
                (\BoxRight,\BoxBottom)
                (\BoxRight,\BoxTop)
                (\BoxLeft,\BoxTop)
                (\BoxLeft,\BoxBottom)
            };
            \noexpand\addplot[
                only marks, mark=#4, mark size=\SeriesMarkSize,
                color=#3,
                mark options={
                    draw=#3,
                    fill=\SeriesFill,
                    line width=0.3pt
                },
                forget plot
            ] coordinates {\SeriesCoords};
        }%
        \RenderSeries
    \fi
}

\newcommand{\PlotApplication}[1]{%
    \PlotSeriesWithBox{#1}{HPA}
        {hpaColor}{*}%
    \PlotSeriesWithBox{#1}{Erlang}
        {erlangColor}{triangle*}%
    \PlotSeriesWithBox{#1}{opusLow}
        {opusColor}{diamond}%
    \PlotSeriesWithBox{#1}{opusHigh}
        {opusColor}{diamond*}%
    \PlotSeriesWithBox{#1}{jevSingle}
        {jevColor}{square}%
    \PlotSeriesWithBox{#1}{jevBatch}
        {jevColor}{square*}%
    \PlotSeriesWithBox{#1}{agar}
        {agarColor}{pentagon*}%

    % Draw Laya last so the plus markers remain visible.
    \PlotSeriesWithBox{#1}{Laya}
        {layaColor}{filled plus}%
}

\begin{figure}[!t]
\centering
\begin{tikzpicture}
\begin{groupplot}[
    group style={group size=1 by 3,vertical sep=1.45cm},
    scale only axis,
    width=0.8\columnwidth,
    height=3.175cm,
    xmin=5, xmax=420,
    ymin=0, ymax=350,
    xtick={0,100,200,300,400},
    ytick={0,100,200,300},
    xlabel={Replicas},
    ylabel={Median p95 latency (ms)},
    label style={font=\small},
    tick label style={font=\footnotesize},
    tick align=inside,
    xtick pos=bottom,
    ytick pos=left,
    major tick length=3pt,
    tick style={draw=black!35,line width=0.6pt},
    axis line style={draw=black!35,line width=0.5pt},
    grid=major,
    grid style={draw=black!15,line width=0.5pt},
    extra y ticks={300},
    extra y tick labels={},
    extra y tick style={
        grid=major,
        grid style={
            draw=red, line width=0.9pt,
            dash pattern=on 3.6pt off 2.7pt
        }
    },
    set layers=standard,
    clip mode=individual,
    clip marker paths=true
]

\nextgroupplot[
    xmin=20, xmax=60,
    xtick={20,40,60},
    xlabel style={
        at={(axis description cs:0.5,0)},
        anchor=north,
        yshift=0pt,
        inner sep=0pt
    }
]
\PlotApplication{OB}

\pgfplotsset{
    legend columns=4,
    legend cell align=left,
    legend style={
        at={(0.5,1.10)},
        anchor=south,
        font=\scriptsize,
        draw=black!25,
        fill=white,
        inner xsep=3pt,
        inner ysep=3pt,
        row sep=2pt,
        /tikz/every even column/.append style={
            column sep=3pt
        }
    }
}

% Autoscaler legend.
\addlegendimage{
    only marks,
    mark=*,
    mark size=2.7pt,
    mark options={solid,draw=hpaColor,fill=hpaColor, yshift=0.2pt}
}
\addlegendentry{HPA}

\addlegendimage{
    only marks,
    mark=triangle*,
    mark size=2.7pt,
    mark options={solid,draw=erlangColor,fill=erlangColor, yshift=0.2pt}
}
\addlegendentry{Erlang}

\addlegendimage{
    only marks,
    mark=pentagon*,
    mark size=2.7pt,
    mark options={solid,draw=agarColor,fill=agarColor,yshift=0.2pt}
}
\addlegendentry{Agar}

\addlegendimage{
    only marks,
    mark=filled plus,
    mark size=2.7pt,
    mark options={solid,draw=layaColor,fill=layaColor, yshift=0.2pt}
}
\addlegendentry{Laya}

\addlegendimage{
    only marks,
    mark=diamond,
    mark size=2.7pt,
    mark options={solid,draw=opusColor,fill=white, yshift=0.2pt}
}
\addlegendentry{Opus-Low}

\addlegendimage{
    only marks,
    mark=diamond*,
    mark size=2.7pt,
    mark options={solid,draw=opusColor,fill=opusColor, yshift=0.2pt}
}
\addlegendentry{Opus-High}

\addlegendimage{
    only marks,
    mark=square,
    mark size=2.7pt,
    mark options={solid,draw=jevColor,fill=white,yshift=0.2pt}
}
\addlegendentry{Jev-Single}

\addlegendimage{
    only marks,
    mark=square*,
    mark size=2.7pt,
    mark options={solid,draw=jevColor,fill=jevColor,yshift=0.2pt}
}
\addlegendentry{Jev-Batch}

\nextgroupplot[
    xmin=20, xmax=450,
    xtick={20,40,60,400},
    xticklabels={20,40,60,400},
    x coord trafo/.code={%
        \pgfkeys{/pgf/fpu=true}%
        \pgfmathparse{#1 <= 60 ? #1 : 60+(#1-60)*0.02}%
        \pgfmathfloattofixed{\pgfmathresult}%
        \pgfkeys{/pgf/fpu=false}%
    },
    x coord inv trafo/.code={%
        \pgfkeys{/pgf/fpu=true}%
        \pgfmathparse{#1 <= 60 ? #1 : 60+(#1-60)/0.02}%
        \pgfmathfloattofixed{\pgfmathresult}%
        \pgfkeys{/pgf/fpu=false}%
    },
    xlabel style={
        at={(axis description cs:0.5,0)},
        anchor=north,
        yshift=0pt,
        inner sep=0pt
    }
]
\PlotApplication{SN}

% Mask the x-axis.
\draw[white, line width=3pt]
    ([xshift=9pt]axis cs:60,0)
    ++(-5pt,0) -- ++(12pt,0);

% Smaller zigzag, shifted right of the 60 tick.
\draw[black, line width=0.5pt, line join=round]
    ([xshift=9pt]axis cs:60,0)
    ++(-5pt,0pt)
    -- ++(1pt,0pt)
    -- ++(3pt,-4pt)
    -- ++(3pt,8pt)
    -- ++(3pt,-4pt)
    -- ++(2pt,0pt);

\nextgroupplot[
    xmin=0, xmax=363,
    xtick={0,20,40,60,80,220,240,340,360},
    xticklabels={0,20,40,60,80,220,240,340,360},
    scaled x ticks=false,
    plot coordinates/math parser=false,
    x coord trafo/.code={%
        \pgfmathparse{ifthenelse(#1<=80,#1/20,
            ifthenelse(#1<=220,4+(#1-80)/140,
            ifthenelse(#1<=240,5+(#1-220)/20,
            ifthenelse(#1<=340,6+(#1-240)/100,
            7+(#1-340)/20))))}%
    },
    x coord inv trafo/.code={%
        \pgfmathparse{ifthenelse(#1<=4,#1*20,
            ifthenelse(#1<=5,80+(#1-4)*140,
            ifthenelse(#1<=6,220+(#1-5)*20,
            ifthenelse(#1<=7,240+(#1-6)*100,
            340+(#1-7)*20))))}%
    },
    ymax=90000,
    ytick={0,100,200,300,60000},
    yticklabels={0,100,200,300,60K},
    y coord trafo/.code={%
        \pgfkeys{/pgf/fpu=true}%
        \pgfmathparse{#1 <= 350 ? #1 : 350+(#1-350)*0.001}%
        \pgfmathfloattofixed{\pgfmathresult}%
        \pgfkeys{/pgf/fpu=false}%
    },
    y coord inv trafo/.code={%
        \pgfkeys{/pgf/fpu=true}%
        \pgfmathparse{#1 <= 350 ? #1 : 350+(#1-350)*1000}%
        \pgfmathfloattofixed{\pgfmathresult}%
        \pgfkeys{/pgf/fpu=false}%
    },
]
\PlotApplication{TS}
% Mask the underlying y-axis.
\draw[white, line width=2pt]
    (rel axis cs:0,0.8)
    ++(0pt,-6pt) -- ++(0pt,16pt);

% Zigzag with short vertical connections.
\draw[black, line width=0.5pt, line join=round]
    (rel axis cs:0,0.8)
    ++(0pt,-7pt)
    -- ++(0pt,1pt)
    -- ++(-4pt,4pt)
    -- ++(6pt,4pt)
    -- ++(-6pt,4pt)
    -- ++(4pt,4pt)
    -- ++(0pt,1pt);

% First break: 80–200.
\draw[white, line width=3pt]
    ([xshift=5pt]axis cs:90,0)
    ++(-5pt,0) -- ++(12pt,0);

\draw[black, line width=0.5pt, line join=round]
    ([xshift=5pt]axis cs:90,0)
    ++(-5pt,0pt)
    -- ++(3pt,0pt)
    -- ++(3pt,-4pt)
    -- ++(3pt,8pt)
    -- ++(3pt,-4pt);

% Second break: 250–400.
\draw[white, line width=3pt]
    ([xshift=1pt]axis cs:280,0)
    ++(-5pt,0) -- ++(12pt,0);

\draw[black, line width=0.5pt, line join=round]
    ([xshift=1pt]axis cs:280,0)
    ++(-5pt,0pt)
    -- ++(3pt,0pt)
    -- ++(3pt,-4pt)
    -- ++(3pt,8pt)
    -- ++(3pt,-4pt);

\end{groupplot}

% Online Boutique label: just above the Social Network graph.
\node[
    font=\small\bfseries,
    anchor=south,
    inner sep=0pt,
    outer sep=0pt
]
    at ([yshift=6pt]group c1r2.north)
    {(a) Online Boutique};

% Social Network label: just above the TeaStore graph.
\node[
    font=\small\bfseries,
    anchor=south,
    inner sep=0pt,
    outer sep=0pt
]
    at ([yshift=6pt]group c1r3.north)
    {(b) Social Network};

% TeaStore label.
\node[font=\small\bfseries]
    at ([yshift=-35pt]group c1r3.south)
    {(c) TeaStore};

\end{tikzpicture}

\caption{Per-endpoint median p95 latency when horizontally scaling replicas using HPA, Erlang, Agar, Claude Opus 5.5 (low and high reasoning effort), Laya, and Jev on CloudLab. Jev-Single sends a separate request per service every 15\,s, while Jev-Batched sends one batched request for all services every 15\,s. The horizontal dashed line marks the 300\,ms \textcolor{red!70!black}{SLO}. For each application, the boundary boxes show average running replicas, except for Jev-Single on Social Network and TeaStore and Laya on TeaStore, where they show requested replicas; excess pods remain Pending due to cluster capacity limits.}
\label{fig:autoscaler-latencies-cloudlab}

\end{figure}
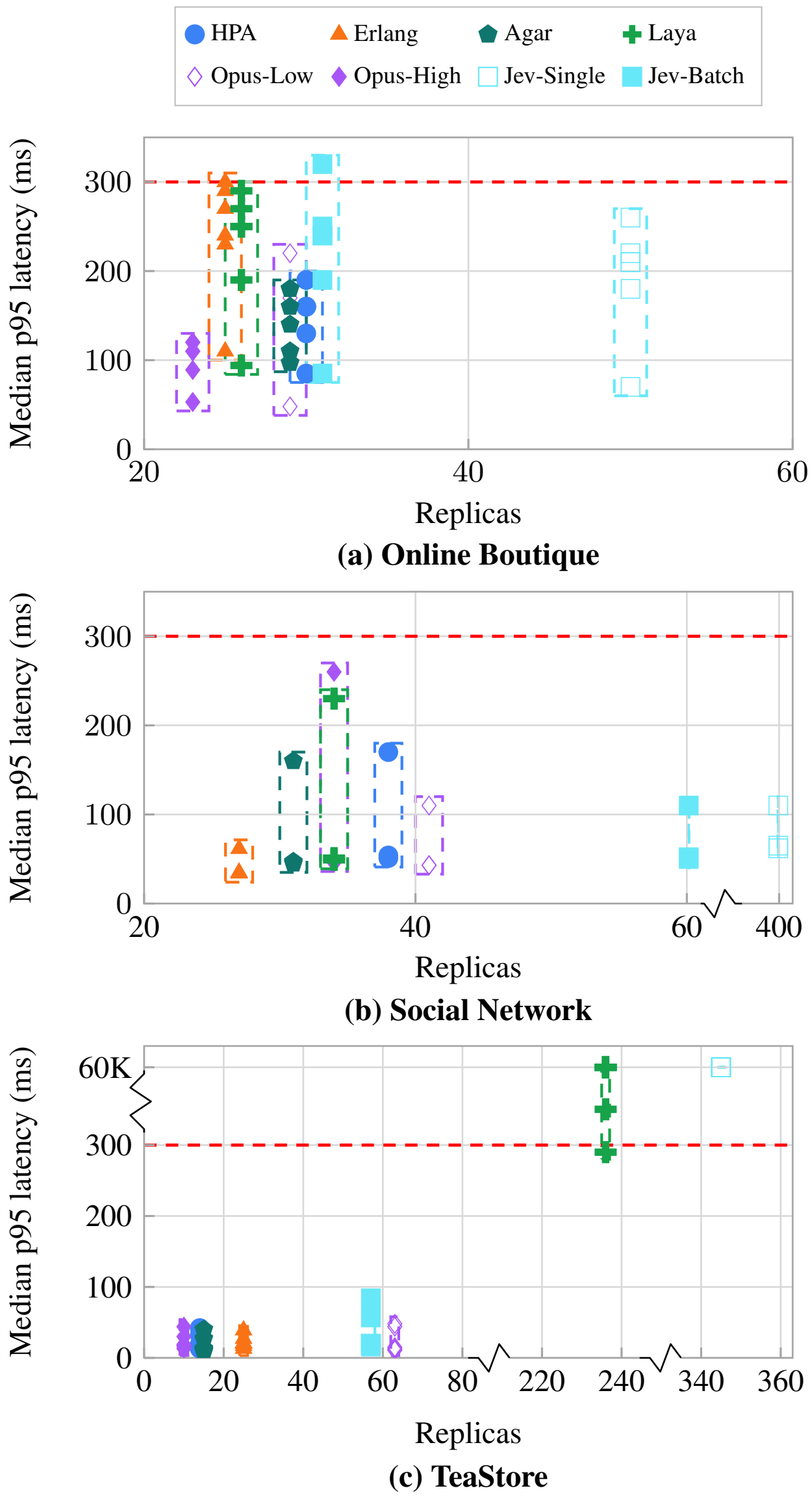

%% file: eval_autoscaling_results.tex
\subsubsection{Results}

Figure~\ref{fig:autoscaler-latencies-cloudlab} compares the resource allocations and per-endpoint latencies achieved by each autoscaler. Additional reasoning effort substantially improves Opus 5.5's decisions, particularly on TeaStore, where Opus-High achieves the lowest resource allocation among all autoscalers. However, greater model sophistication does not consistently yield better results: on Social Network, Laya achieves lower worst-endpoint latency than Opus-High with approximately the same number of replicas.

The information available to the models also matters. Although Laya's requests are batched for inference efficiency, each decision sees only its own service's resource state. On TeaStore, Laya and Jev-Single request more replicas than the cluster can accommodate, leaving excess pods in the Pending state while violating the SLO. Providing Jev with application-wide state reduces replica allocation by over 6$\times$ on both TeaStore and Social Network, and eliminates the SLO violations on TeaStore. However, Jev-Batched remains less resource-efficient than the other autoscalers on Social Network.

%% file: eval_dram.tex
\subsection{Dynamic DRAM Sizing}\label{subsec:dram-eval}

We evaluate how much DRAM the \name sizers can save while maintaining at least 95\% of the performance achieved with static DRAM provisioning.

\input{eval_dram_setup}

\input{figures/dram-sizing-eval}

\input{eval_dram_results}

%% file: eval_dram_setup.tex
\subsubsection{Experimental Setup}

\paragraph{Hardware Setup.}
We run our experiments on a \texttt{c220g5} CloudLab~\cite{cloudlab} node running Linux v6.18. We use remote NUMA memory to emulate the slow CXL tier. We lower the uncore frequency of the remote node to emulate longer slow tier latencies. Unloaded latency is $\approx$90\,ns for local DRAM and $\approx$180\,ns for the emulated CXL tier.

\begin{table}[t]
\scriptsize
\centering
\setlength{\abovecaptionskip}{-2pt}
\setlength{\belowcaptionskip}{0pt}
%\rev{
\resizebox{\linewidth}{!}{
\begin{tabular}{@{}lllll@{}}
\toprule
\textbf{Workload} & \textbf{Inputs} & \textbf{RSS} & \textbf{Description} \\ \midrule
Liblinear~\cite{fan2008liblinear} & kddb & 20.69 & Logistic regression-based classifier. \\ \midrule
% GapBS-BC~\cite{beamer2015gap} & twitter  & 13.08 & Compute the measure of centrality  \\\midrule
GapBS-PR~\cite{beamer2015gap} & twitter  & 12.32 & Compute PageRank score of a graph. \\
\midrule
GapBS-CC~\cite{beamer2015gap} & kronecker  & 69.29 & Compute connected components of graph \\ \midrule
\multirow{2}{*}{Silo~\cite{tu2013speedy}} & TPC-C & 75.68 & \multirow{2}{*}{In-memory transactional database.} \\
 & YCSB-C & 71.40 & \\ \midrule
Btree~\cite{reto2020btree} & - & 42.01 &  In-memory index lookup benchmark. \\ \midrule
XSBench~\cite{tramm2014xsbench} & - & 64.97 & Compute kernel of MCNP algorithm. \\ \midrule
GUPS ~\cite{plimpton2006simple} & 8 GiB hot & 64.03 & Skewed accesses with dynamic hotset. \\
\bottomrule
\end{tabular}
}
%}
\caption{DRAM sizing workload characteristics. Resident set size (RSS) is in GiB.}
\label{tab:benchmark_specs}
% \vspace{-4ex}
\end{table}

\paragraph{Workloads.}
For our evaluation, we select a set of eight representative and diverse workloads shown in Table~\ref{tab:benchmark_specs}. These workloads have been widely used to evaluate tiered memory systems in prior works~\cite{hemem2021, memtis2023, johnnyCache2023,autotiering2017,tiersizer}. We run these workloads with 16 threads, large enough to saturate the memory bandwidth of each system.

\paragraph{Baselines.} We compare the \name DRAM sizers against TierSizer~\cite{tiersizer}, a DRAM sizer that uses analytical models to predict application slowdown at unexecuted DRAM sizes and adjust allocations at runtime, and a static 2:1 DRAM:CXL ratio, representative of current state-of-the-art deployments~\cite{tpp2023}. Pond~\cite{cxlpond} predicts DRAM allocations at VM startup and reverts to fully local DRAM if performance degrades, whereas our approach continuously adjusts DRAM allocations at runtime. We set the performance target to a 5\% slowdown relative to this 2:1 baseline.

%% file: figures/dram-sizing-eval.tex
\definecolor{tiersizerColor}{HTML}{F97316}
\definecolor{layaColor}{HTML}{16A34A}
\definecolor{opusColor}{HTML}{A855F7}
\definecolor{jevColor}{HTML}{67E8F9}

\pgfplotstableread[
    col sep=comma,
    header=true
]{data/dram-sizing-eval.csv}\dramdata

% #1 application
% #2 sizer
% #3 column
% #4 output macro
\newcommand{\GetDRAMValue}[4]{%
    \gdef#4{0}%
    \pgfplotstableforeachcolumnelement{app}\of\dramdata
        \as\CurrentApp{%
        \pgfplotstablegetelem
            {\pgfplotstablerow}{sizer}\of\dramdata
        \edef\CurrentSizer{\pgfplotsretval}%
        \ifdefstring{\CurrentApp}{#1}{%
            \ifdefstring{\CurrentSizer}{#2}{%
                \pgfplotstablegetelem
                    {\pgfplotstablerow}{#3}\of\dramdata
                \xdef#4{\pgfplotsretval}%
            }{}%
        }{}%
    }%
}

% #1 application
% #2 x position
\newcommand{\PlotDRAMApplication}[2]{%
    \GetDRAMValue{#1}{static}{avg_dram_size}{\StaticDRAM}%
    \GetDRAMValue{#1}{static}{avg_perf}{\StaticPerf}%

    \GetDRAMValue{#1}{tiersizer}{avg_dram_size}{\TierAvg}%
    \GetDRAMValue{#1}{tiersizer}{min_dram_size}{\TierMin}%
    \GetDRAMValue{#1}{tiersizer}{avg_perf}{\TierPerf}%

    \GetDRAMValue{#1}{laya}{avg_dram_size}{\LayaAvg}%
    \GetDRAMValue{#1}{laya}{min_dram_size}{\LayaMin}%
    \GetDRAMValue{#1}{laya}{avg_perf}{\LayaPerf}%

    \GetDRAMValue{#1}{opusLow}{avg_dram_size}{\OpusLowAvg}%
    \GetDRAMValue{#1}{opusLow}{min_dram_size}{\OpusLowMin}%
    \GetDRAMValue{#1}{opusLow}{avg_perf}{\OpusLowPerf}%
    
    \GetDRAMValue{#1}{opusHigh}{avg_dram_size}{\OpusHighAvg}%
    \GetDRAMValue{#1}{opusHigh}{min_dram_size}{\OpusHighMin}%
    \GetDRAMValue{#1}{opusHigh}{avg_perf}{\OpusHighPerf}%

    \GetDRAMValue{#1}{jev}{avg_dram_size}{\JevAvg}%
    \GetDRAMValue{#1}{jev}{min_dram_size}{\JevMin}%
    \GetDRAMValue{#1}{jev}{avg_perf}{\JevPerf}%

    % Normalized DRAM allocation.
    \pgfmathsetmacro{\TierAvgNorm}{\TierAvg/\StaticDRAM}%
    \pgfmathsetmacro{\TierMinNorm}{\TierMin/\StaticDRAM}%
    \pgfmathsetmacro{\LayaAvgNorm}{\LayaAvg/\StaticDRAM}%
    \pgfmathsetmacro{\LayaMinNorm}{\LayaMin/\StaticDRAM}%

    % Percentage savings.
    \pgfmathtruncatemacro{\TierSaving}
        {round(100*(1-\TierAvgNorm))}%
    \pgfmathtruncatemacro{\LayaSaving}
        {round(100*(1-\LayaAvgNorm))}%

    % Normalized performance (inverse execution time).
    \pgfmathsetmacro{\TierPerfNorm}
        {\StaticPerf/\TierPerf}%
    \pgfmathsetmacro{\LayaPerfNorm}
        {\StaticPerf/\LayaPerf}%

    \pgfmathsetmacro{\OpusLowAvgNorm}{\OpusLowAvg/\StaticDRAM}%
    \pgfmathsetmacro{\OpusLowMinNorm}{\OpusLowMin/\StaticDRAM}%
    \pgfmathsetmacro{\OpusHighAvgNorm}{\OpusHighAvg/\StaticDRAM}%
    \pgfmathsetmacro{\OpusHighMinNorm}{\OpusHighMin/\StaticDRAM}%
    \pgfmathsetmacro{\JevAvgNorm}{\JevAvg/\StaticDRAM}%
    \pgfmathsetmacro{\JevMinNorm}{\JevMin/\StaticDRAM}%

    \pgfmathtruncatemacro{\OpusLowSaving}
        {round(100*(1-\OpusLowAvgNorm))}%
    \pgfmathtruncatemacro{\OpusHighSaving}
        {round(100*(1-\OpusHighAvgNorm))}%
    \pgfmathtruncatemacro{\JevSaving}
        {round(100*(1-\JevAvgNorm))}%

    \pgfmathsetmacro{\OpusLowPerfNorm}{\StaticPerf/\OpusLowPerf}%
    \pgfmathsetmacro{\OpusHighPerfNorm}{\StaticPerf/\OpusHighPerf}%
    \pgfmathsetmacro{\JevPerfNorm}{\StaticPerf/\JevPerf}%

    \ifnum\DRAMPanel=1
        % Average DRAM bars.
        \edef\RenderBars{%
            \noexpand\addplot[
                ybar, bar width=3pt, bar shift=0pt,
                fill=tiersizerColor, draw=none, forget plot
            ] coordinates {(#2-0.32,\TierAvgNorm)};
        
            \noexpand\addplot[
                ybar, bar width=3pt, bar shift=0pt,
                fill=layaColor, draw=none, forget plot
            ] coordinates {(#2-0.16,\LayaAvgNorm)};
        
            \noexpand\addplot[
                ybar, bar width=3pt, bar shift=0pt,
                fill=white, draw=opusColor,
                pattern={Lines[
                    angle=45,
                    distance=3pt,
                    line width=0.4pt
                ]},
                pattern color=opusColor,
                forget plot
            ] coordinates {(#2,\OpusLowAvgNorm)};
        
            \noexpand\addplot[
                ybar, bar width=3pt, bar shift=0pt,
                fill=opusColor, draw=none, forget plot
            ] coordinates {(#2+0.16,\OpusHighAvgNorm)};
        
            \noexpand\addplot[
                ybar, bar width=3pt, bar shift=0pt,
                fill=jevColor, draw=none, forget plot
            ] coordinates {(#2+0.32,\JevAvgNorm)};
        }%
        \RenderBars

        % Minimum DRAM markers spanning the full bar width.
        \edef\RenderMinMarkers{%
            \noexpand\draw[black, line width=1pt]
                ([xshift=-1.5pt]axis cs:#2-0.32,\TierMinNorm) --
                ([xshift=1.5pt]axis cs:#2-0.32,\TierMinNorm);
        
            \noexpand\draw[black, line width=1pt]
                ([xshift=-1.5pt]axis cs:#2-0.16,\LayaMinNorm) --
                ([xshift=1.5pt]axis cs:#2-0.16,\LayaMinNorm);
        
            \noexpand\draw[black, line width=1pt]
                ([xshift=-1.5pt]axis cs:#2,\OpusLowMinNorm) --
                ([xshift=1.5pt]axis cs:#2,\OpusLowMinNorm);
        
            \noexpand\draw[black, line width=1pt]
                ([xshift=-1.5pt]axis cs:#2+0.16,\OpusHighMinNorm) --
                ([xshift=1.5pt]axis cs:#2+0.16,\OpusHighMinNorm);
        
            \noexpand\draw[black, line width=1pt]
                ([xshift=-1.5pt]axis cs:#2+0.32,\JevMinNorm) --
                ([xshift=1.5pt]axis cs:#2+0.32,\JevMinNorm);
        }%
        \RenderMinMarkers

    \else
        % Performance bars.
        \edef\RenderPerf{%
            \noexpand\addplot[
                ybar, bar width=3pt, bar shift=0pt,
                fill=tiersizerColor, draw=none, forget plot
            ] coordinates {(#2-0.32,\TierPerfNorm)};
        
            \noexpand\addplot[
                ybar, bar width=3pt, bar shift=0pt,
                fill=layaColor, draw=none, forget plot
            ] coordinates {(#2-0.16,\LayaPerfNorm)};
        
            \noexpand\addplot[
                ybar, bar width=3pt, bar shift=0pt,
                fill=white, draw=opusColor,
                pattern={Lines[
                    angle=45,
                    distance=3pt,
                    line width=0.4pt
                ]},
                pattern color=opusColor,
                forget plot
            ] coordinates {(#2,\OpusLowPerfNorm)};
        
            \noexpand\addplot[
                ybar, bar width=3pt, bar shift=0pt,
                fill=opusColor, draw=none, forget plot
            ] coordinates {(#2+0.16,\OpusHighPerfNorm)};
        
            \noexpand\addplot[
                ybar, bar width=3pt, bar shift=0pt,
                fill=jevColor, draw=none, forget plot
            ] coordinates {(#2+0.32,\JevPerfNorm)};
        }%
        \RenderPerf
    \fi
}

\newcommand{\PlotAllDRAMApplications}{%
    \PlotDRAMApplication{liblinear}{1}%
    % \PlotDRAMApplication{bc-twitter}{2}%
    \PlotDRAMApplication{pr-twitter}{2}%
    \PlotDRAMApplication{cc-twitter}{3}%
    \PlotDRAMApplication{silo-tpcc}{4}%
    \PlotDRAMApplication{silo-ycsb}{5}%
    \PlotDRAMApplication{btree}{6}%
    \PlotDRAMApplication{xsbench}{7}%
    \PlotDRAMApplication{gups}{8}%
}

\begin{figure}[!t]
\centering

\begin{tikzpicture}
\begin{groupplot}[
    group style={
        group size=1 by 2,
        vertical sep=2.5cm
    },
    scale only axis,
    width=0.75\columnwidth,
    height=3.2cm,
    xmin=0.5,
    xmax=8.5,
    xtick={1,2,3,4,5,6,7,8},
    xticklabels={
        Liblinear,
        % BC,
        PR,
        CC,
        \shortstack{Silo-\\TPC-C},
        \shortstack{Silo-\\YCSB},
        Btree,
        XSBench,
        GUPS
    },
    xticklabel style={
        rotate=45,
        anchor=north east,
        font=\tiny
    },
    label style={font=\small},
    tick align=inside,
    xtick pos=bottom,
    ytick pos=left,
    major tick length=3pt,
    tick style={
        draw=black!35,
        line width=0.6pt
    },
    axis line style={
        draw=black!35,
        line width=0.5pt
    },
    ymajorgrids=true,
    grid style={
        draw=black!15,
        line width=0.5pt
    },
    set layers=standard,
    clip=false
]

% --------------------------------------------------
% (a) DRAM savings
% --------------------------------------------------

\nextgroupplot[
    ymin=0,
    ymax=1.15,
    ytick={0,0.5,1.0},
    ylabel={
        DRAM\\
        (norm. to static)
    },
    ylabel style={align=center},
    legend columns=4,
    legend cell align=left,
    legend image post style={xscale=0.7},
    legend style={
        at={(0.5,1.08)},
        anchor=south,
        font=\scriptsize,
        draw=black!25,
        fill=none,
        column sep=2pt,
        /tikz/every even column/.append style={
            column sep=2pt
        }
    }
]

\def\DRAMPanel{1}

% Static baseline.
\addplot[
    black,
    densely dotted,
    line width=0.9pt,
    forget plot
] coordinates {
    (0.5,1)
    (8.5,1)
};

\PlotAllDRAMApplications

% Legend (two rows).
\addlegendimage{
    line legend,
    black,
    densely dotted,
    line width=0.9pt
}
\addlegendentry{Static 2:1}

\addlegendimage{area legend, fill=tiersizerColor, draw=none}
\addlegendentry{TierSizer}

\addlegendimage{area legend, fill=layaColor, draw=none}
\addlegendentry{Laya}

\addlegendimage{area legend, fill=jevColor, draw=none}
\addlegendentry{Jev}

\addlegendimage{
    line legend,
    black,
    line width=1pt
}
\addlegendentry{Peak Savings}

\addlegendimage{
    area legend,
    fill=white,
    draw=opusColor,
    pattern={Lines[
        angle=45,
        distance=3pt,
        line width=0.4pt
    ]},
    pattern color=opusColor
}
\addlegendentry{Opus-Low}

\addlegendimage{area legend, fill=opusColor, draw=none}
\addlegendentry{Opus-High}

% --------------------------------------------------
% (b) Performance
% --------------------------------------------------

\nextgroupplot[
    ymin=0.6,
    ymax=1.06,
    ytick={0.6,0.8,1.0},
    extra y ticks={0.95},
    extra y tick labels={\textcolor{red}{0.95}},
    ylabel={
        Normalized\\
        performance
    },
    ylabel style={align=center},
    legend columns=1,
    legend style={
        at={(0.5,1.08)},
        anchor=south,
        font=\scriptsize,
        draw=black!50,
        fill=white,
        line width=0.5pt
    }
]

\def\DRAMPanel{2}

% Static baseline.
\addplot[
    black,
    densely dotted,
    line width=0.9pt,
    forget plot
] coordinates {
    (0.5,1)
    (8.5,1)
};

\PlotAllDRAMApplications

% Target performance (drawn on top of bars).
\addplot[
    red,
    dashed,
    line width=0.9pt,
    forget plot
] coordinates {
    (0.5,0.95)
    (8.5,0.95)
};

% Legend.
\addlegendimage{
    line legend,
    red,
    dashed,
    line width=0.9pt
}
\addlegendentry{Target performance}

\end{groupplot}

% Panel labels.
\node[
    font=\small,
    anchor=north
] at ([yshift=-28pt]group c1r1.south)
    {(a) DRAM savings.};

\node[
    font=\small,
    anchor=north
] at ([yshift=-28pt]group c1r2.south)
    {(b) Performance.};

\end{tikzpicture}

\caption{
DRAM usage and performance achieved with TierSizer, Laya, Opus 5.5 (low and high reasoning effort), and Jev, normalized to static 2:1 DRAM provisioning. Bars show average DRAM usage and performance, while black markers indicate minimum DRAM usage (peak savings). The dashed red line marks the 95\% performance target.
}
\label{fig:dram-sizing-eval}

\end{figure}
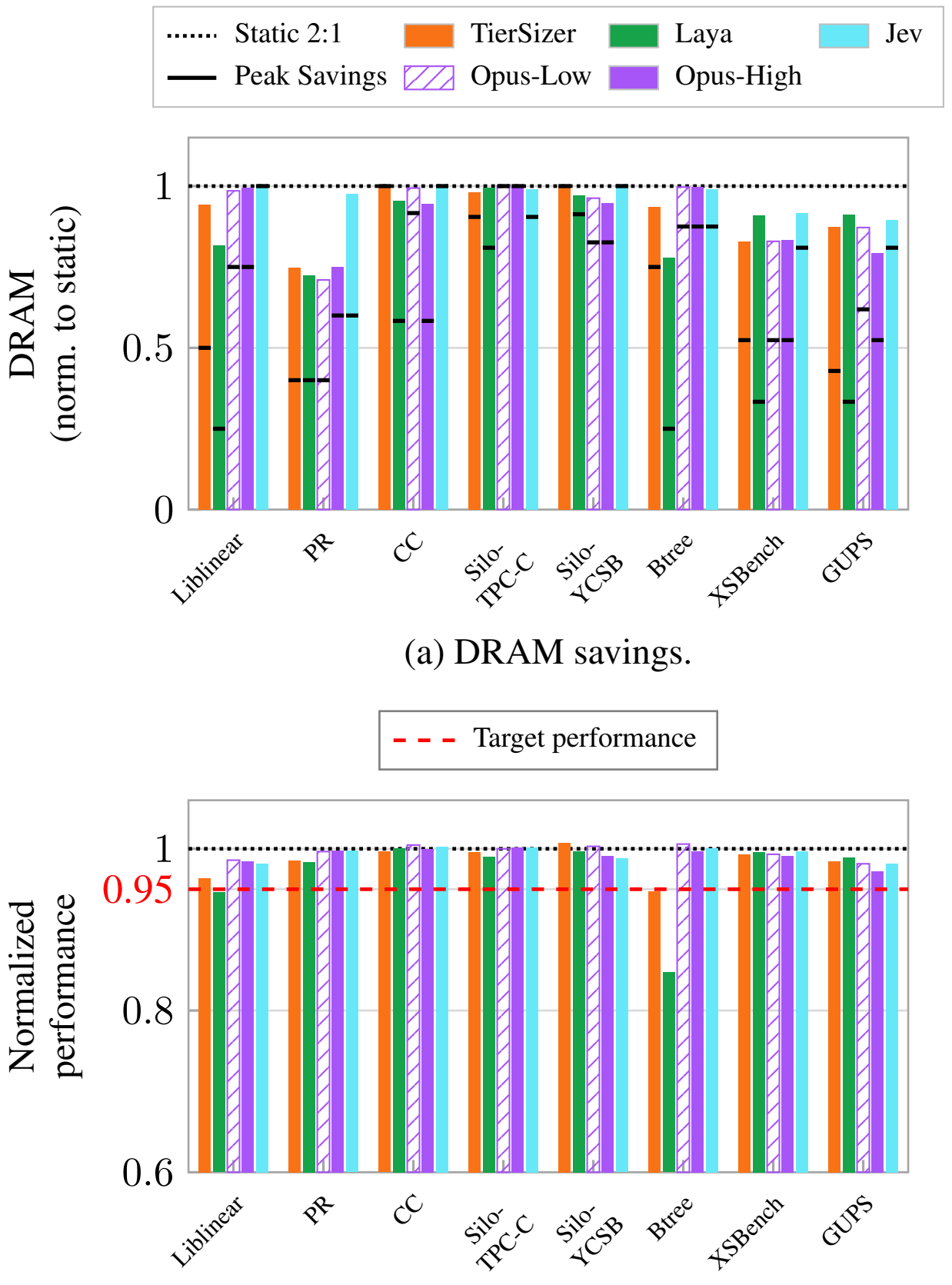

%% file: eval_dram_results.tex
\subsubsection{Results}

Figure~\ref{fig:dram-sizing-eval} compares DRAM allocations and performance across the eight workloads. The \name sizers achieve savings comparable to TierSizer on several workloads without learning application-specific performance models. On PageRank, Laya and Opus 5.5 reduce average DRAM allocation by approximately 25--30\% while meeting the 95\% performance target. Opus 5.5 also achieves savings of approximately 15--20\% on XSBench and GUPS while maintaining performance. On both Silo workloads, all methods retain allocations close to static provisioning.

The models differ in how aggressively they reduce DRAM. Jev generally retains more DRAM, particularly on Liblinear, PageRank, and connected components. Laya is more aggressive, achieving substantial savings on Liblinear and BTree, but violates the performance target on BTree, where performance falls to approximately 85\% of the static baseline. In contrast to autoscaling, Opus-Low and Opus-High achieve nearly identical DRAM allocations and performance across most workloads, with little benefit from additional reasoning effort.

%% file: eval_ffmpeg.tex
\subsection{Video Compression}

\input{figures/ffmpeg-eval}

\input{eval_ffmpeg_setup}

\input{figures/ffmpeg-eval-ablation}

\input{eval_ffmpeg_results}

\input{eval_ffmpeg_ablation}

%% file: figures/ffmpeg-eval.tex
\definecolor{defaultColor}{HTML}{F97316}
\definecolor{layaColor}{HTML}{16A34A}
\definecolor{opusColor}{HTML}{A855F7}
\definecolor{jevColor}{HTML}{67E8F9}

\pgfplotstableread[
    col sep=comma,
    header=true
]{data/ffmpeg-eval.csv}\videodata

% #1 clip, #2 encoder, #3 color, #4 marker, #5 y-column
\newcommand{\PlotVideoTradeoff}[5]{%
    \def\TradeoffCoords{}%
    \pgfplotstableforeachcolumnelement{mode}\of\videodata
        \as\CurrentMode{%
        \pgfplotstablegetelem
            {\pgfplotstablerow}{clip}\of\videodata
        \edef\CurrentClip{\pgfplotsretval}%
        \pgfplotstablegetelem
            {\pgfplotstablerow}{window_s}\of\videodata
        \edef\CurrentWindow{\pgfplotsretval}%
        \ifnum\pdfstrcmp{\CurrentClip}{#1}=0\relax
            \ifnum\pdfstrcmp{\CurrentMode}{#2}=0\relax
                \ifnum\pdfstrcmp{#2}{raw}=0\relax
                    \pgfplotstablegetelem
                        {\pgfplotstablerow}{wall_seconds}\of\videodata
                    \edef\CurrentTime{\pgfplotsretval}%
                    \pgfplotstablegetelem
                        {\pgfplotstablerow}{#5}\of\videodata
                    \edef\CurrentY{\pgfplotsretval}%
                    \xdef\TradeoffCoords{%
                        \TradeoffCoords
                        (\CurrentTime,\CurrentY)%
                    }%
                \else
                    \ifnum\pdfstrcmp{\CurrentWindow}{5}=0\relax
                        \pgfplotstablegetelem
                            {\pgfplotstablerow}{wall_seconds}\of\videodata
                        \edef\CurrentTime{\pgfplotsretval}%
                        \pgfplotstablegetelem
                            {\pgfplotstablerow}{#5}\of\videodata
                        \edef\CurrentY{\pgfplotsretval}%
                        \xdef\TradeoffCoords{%
                            \TradeoffCoords
                            (\CurrentTime,\CurrentY)%
                        }%
                    \fi
                \fi
            \fi
        \fi
    }%
    \ifx\TradeoffCoords\empty
    \else
        \addplot[
            only marks,
            mark=#4,
            mark size=2.5pt,
            mark options={draw=#3,fill=white,line width=0.8pt},
            forget plot
        ] coordinates {\TradeoffCoords};
    \fi
}

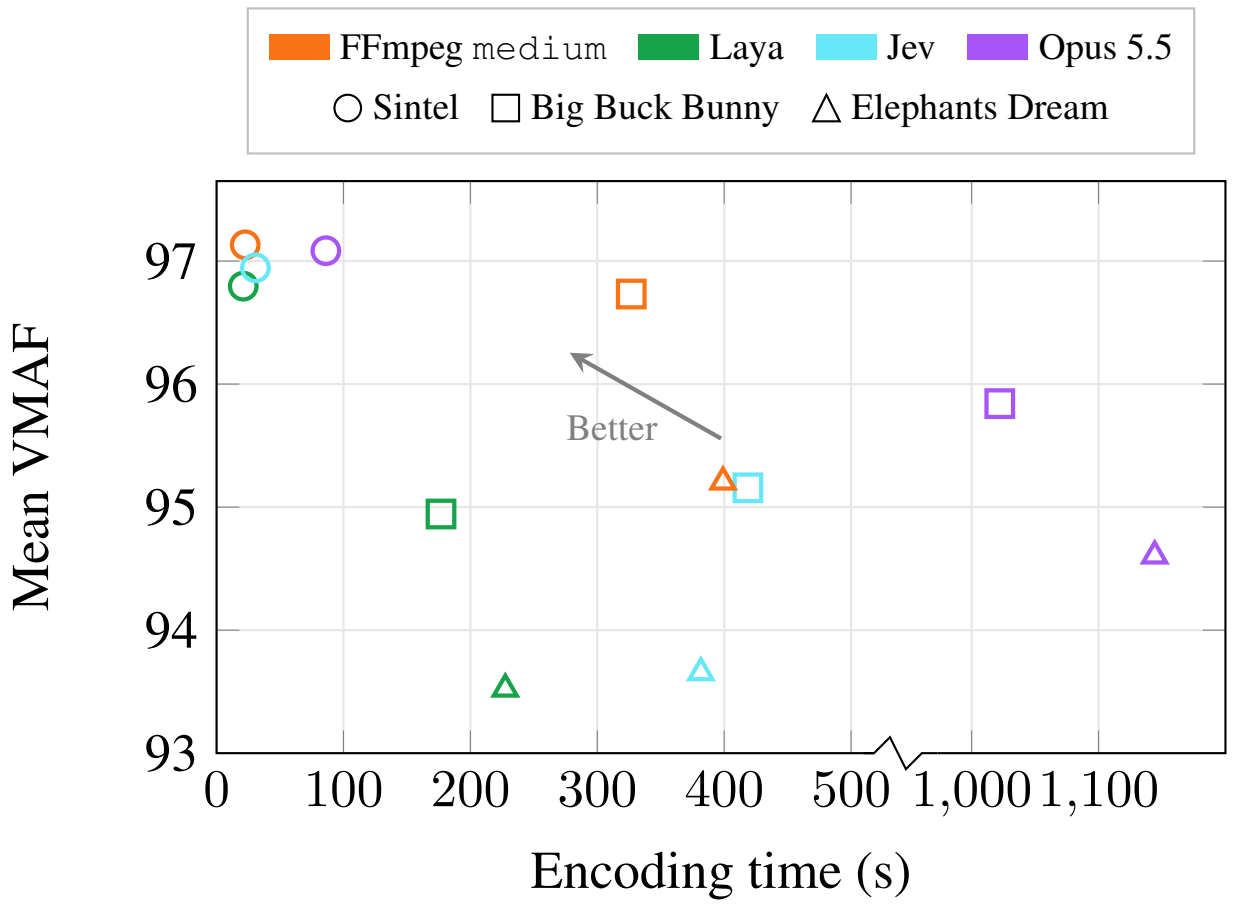
\begin{figure}[!t]
\centering
\begin{tikzpicture}

\begin{axis}[
    width=\columnwidth,
    height=0.65\columnwidth,
    scale only axis=false,
    xlabel={Encoding time (s)},
    ylabel={Mean VMAF},
    xmin=0,
    xmax=1200,
    ymin=93,
    ymax=97.65,
    ytick={93,94,95,96,97},
    yticklabels={93,94,95,96,97},
    xtick={0,100,200,300,400,500,1000,1100},
    x coord trafo/.code={
        \pgfmathparse{#1 > 950 ? #1 - 405 :
                      (#1 > 500 ? 500 + (#1 - 500)*0.1 : #1)}
    },
    x coord inv trafo/.code={
        \pgfmathparse{#1 > 545 ? #1 + 405 :
                      (#1 > 500 ? 500 + (#1 - 500)/0.1 : #1)}
    },
    grid=major,
    grid style={black!10},
    tick align=inside
]

% Sintel: circles
\PlotVideoTradeoff{st}{raw}{defaultColor}{o}{vmaf_mean}
\PlotVideoTradeoff{st}{laya}{layaColor}{o}{vmaf_mean}
\PlotVideoTradeoff{st}{jev}{jevColor}{o}{vmaf_mean}
\PlotVideoTradeoff{st}{opus}{opusColor}{o}{vmaf_mean}

% Big Buck Bunny: squares
\PlotVideoTradeoff{bbb}{raw}{defaultColor}{square}{vmaf_mean}
\PlotVideoTradeoff{bbb}{laya}{layaColor}{square}{vmaf_mean}
\PlotVideoTradeoff{bbb}{jev}{jevColor}{square}{vmaf_mean}
\PlotVideoTradeoff{bbb}{opus}{opusColor}{square}{vmaf_mean}

% Sintel: circles
\PlotVideoTradeoff{ed}{raw}{defaultColor}{triangle}{vmaf_mean}
\PlotVideoTradeoff{ed}{laya}{layaColor}{triangle}{vmaf_mean}
\PlotVideoTradeoff{ed}{jev}{jevColor}{triangle}{vmaf_mean}
\PlotVideoTradeoff{ed}{opus}{opusColor}{triangle}{vmaf_mean}

\draw[
    ->,
    >=stealth,
    thick,
    gray
]
    (rel axis cs:0.50,0.55)
    -- node[midway,below left,xshift=6pt,font=\scriptsize] {Better}
    (rel axis cs:0.35,0.70);

\coordinate (xbreak) at (axis cs:890,93);

\end{axis}

% Mask the axis border
\fill[white]
    ([xshift=-5pt,yshift=-3pt]xbreak)
    rectangle
    ([xshift=7pt,yshift=5pt]xbreak);

% Draw zigzag
\draw[black, line width=0.4pt]
    ([xshift=-7pt]xbreak)
    -- ++(2pt,0pt)
    -- ++(3pt,3pt)
    -- ++(3pt,-6pt)
    -- ++(3pt,3pt)
    -- ++(3pt,0pt);

% Shared legend
\node[
    anchor=south,
    yshift=5pt,
    draw=black!25,
    fill=white,
    inner sep=4pt
] at (current axis.north) {%
    \scriptsize
    \begin{tabular}{@{}c@{}}
        % Encoder legend
        \begin{tabular}{@{}l@{\hspace{2pt}}l
                          @{\hspace{6pt}}l@{\hspace{2pt}}l
                          @{\hspace{6pt}}l@{\hspace{2pt}}l
                          @{\hspace{6pt}}l@{\hspace{2pt}}l@{}}
            \tikz{\fill[defaultColor] (0,-0.07) rectangle (0.4,0.1);}
            & FFmpeg \texttt{medium}
            & \tikz{\fill[layaColor] (0,-0.07) rectangle (0.4,0.1);}
            & Laya
            & \tikz{\fill[jevColor] (0,-0.07) rectangle (0.4,0.1);}
            & Jev
            & \tikz{\fill[opusColor] (0,-0.07) rectangle (0.4,0.1);}
            & Opus 5.5
        \end{tabular}
        \\[3pt]
        % Video legend
        \begin{tabular}{@{}l@{\hspace{2pt}}l
                          @{\hspace{6pt}}l@{\hspace{2pt}}l
                          @{\hspace{6pt}}l@{\hspace{2pt}}l@{}}
            \tikz[baseline=-2pt]{\draw (0,0) circle (2.5pt);}
            & Sintel
            & \tikz[baseline=0.5pt]{\draw (0,0) rectangle (5pt,5pt);}
            & Big Buck Bunny
            & \tikz[baseline=0.5pt]{\draw (0,0) -- (2.5pt,5pt) -- (5pt,0) -- cycle;}
            & Elephants Dream
        \end{tabular}
    \end{tabular}%
};

\end{tikzpicture}

\caption{
Video quality (mean VMAF) versus encoding time across video clips, with encoding parameters updated every 5 seconds. Laya achieves the lowest encoding time on all videos while maintaining quality close to FFmpeg \texttt{medium}. A difference of 6 VMAF points is commonly considered barely perceptible to human viewers~\cite{ozer2017vmaf}. Smaller differences are generally imperceptible.
}
\label{fig:ffmpeg-tradeoff}
\end{figure}

%% file: eval_ffmpeg_setup.tex
\subsubsection{Experimental Setup}

\paragraph{Machine Configuration.} 
We run all encodes on a Dell Precision 3680 with an Intel Core i9-14900 (8 performance and 16 efficiency cores). x264 runs single-threaded through \texttt{ffmpeg}. Opus 5.5 and Jev are accessed through their APIs, while Laya is served locally on an NVIDIA RTX 2000 Ada Generation GPU with 16 GB of memory.

\input{tables/ffmpeg-videos}

\paragraph{Workloads.}
We evaluate three videos from the Xiph.org media collection~\cite{xiph-media}: Big Buck Bunny~\cite{big-buck-bunny}, Elephants Dream~\cite{elephants-dream}, and Sintel Trailer~\cite{sintel}. All videos have a resolution of 720p and a frame rate of 24 frames per second. We encode each video with a target bitrate of 2 Mbps.

\paragraph{Baselines.}
We compare Laya, Jev, and Opus 5.5 against x264’s default configuration using its \texttt{medium} preset. All models select encoding parameters every 5 seconds. We show the impact of varying this decision interval in our ablation experiments (\S\ref{subsec:ffmpeg-ablation}).

%% file: tables/ffmpeg-videos.tex
\begin{table*}[!t]
\centering
\small
\begin{tabular}{@{}ccc@{}}
\toprule
\textbf{Video} & \textbf{Duration} & \textbf{Motion Description} \\
\midrule
Big Buck Bunny~\cite{big-buck-bunny} & 9:56 & Colourful animation, smooth gradients, fine textures, moderate motion \\
Elephants Dream~\cite{elephants-dream} & 10:54 & Dark, highly detailed mechanical scenes with complex camera movement \\
Sintel Trailer~\cite{sintel} & 0:52 & Fast-paced action, rapid cuts and dynamic camera movement \\
\bottomrule
\end{tabular}
\caption{Videos used in our encoding experiments and their motion characteristics. All source videos are 720p at 24 fps, obtained from Xiph.org~\cite{xiph-media}.}
\label{tab:encoding-workloads}
\end{table*}

%% file: figures/ffmpeg-eval-ablation.tex
\definecolor{defaultColor}{HTML}{F97316}
\definecolor{layaColor}{HTML}{16A34A}
\definecolor{opusColor}{HTML}{A855F7}
\definecolor{jevColor}{HTML}{67E8F9}

\pgfplotstableread[col sep=comma]{data/ffmpeg-eval.csv}\videodata

% #1 clip, #2 mode, #3 metric, #4 color
\newcommand{\PlotAblation}[4]{%
    \gdef\AblationCoords{}%
    \gdef\RawBaselineValue{0}%
    \def\VideoMarker{*}%
    \ifnum\pdfstrcmp{#1}{bbb}=0\relax
        \def\VideoMarker{square}%
    \fi
    \ifnum\pdfstrcmp{#1}{ed}=0\relax
        \def\VideoMarker{triangle}%
    \fi
    \pgfplotstableforeachcolumnelement{mode}\of\videodata
        \as\CurrentMode{%
        \pgfplotstablegetelem{\pgfplotstablerow}{clip}\of\videodata
        \edef\CurrentClip{\pgfplotsretval}%
        \ifnum\pdfstrcmp{\CurrentClip}{#1}=0\relax
            \ifnum\pdfstrcmp{\CurrentMode}{#2}=0\relax
                \pgfplotstablegetelem{\pgfplotstablerow}{#3}\of\videodata
                \edef\MetricValue{\pgfplotsretval}%
                \ifnum\pdfstrcmp{#2}{raw}=0\relax
                    \xdef\RawBaselineValue{\MetricValue}%
                \else
                    \pgfplotstablegetelem
                        {\pgfplotstablerow}{window_s}\of\videodata
                    \edef\WindowValue{\pgfplotsretval}%
                    \xdef\AblationCoords{%
                        \AblationCoords(\WindowValue,\MetricValue)%
                    }%
                \fi
            \fi
        \fi
    }%
    \ifnum\pdfstrcmp{#2}{raw}=0\relax
        \addplot[
            color=#4,dashed,thick,mark=none,forget plot
        ] coordinates {(0,\RawBaselineValue)(11,\RawBaselineValue)};
    \else
        \ifnum\pdfstrcmp{#1}{bbb}=0\relax
            \addplot[
                color=#4,mark=square,mark size=2pt,
                thick,forget plot
            ] coordinates {\AblationCoords};
        \else
            \ifnum\pdfstrcmp{#1}{st}=0\relax
                \addplot[
                    color=#4,mark=o,mark size=2pt,
                    thick,forget plot
                ] coordinates {\AblationCoords};
            \else
                \addplot[
                    color=#4,mark=triangle,mark size=2pt,
                    thick,forget plot
                ] coordinates {\AblationCoords};
            \fi
        \fi
    \fi
}

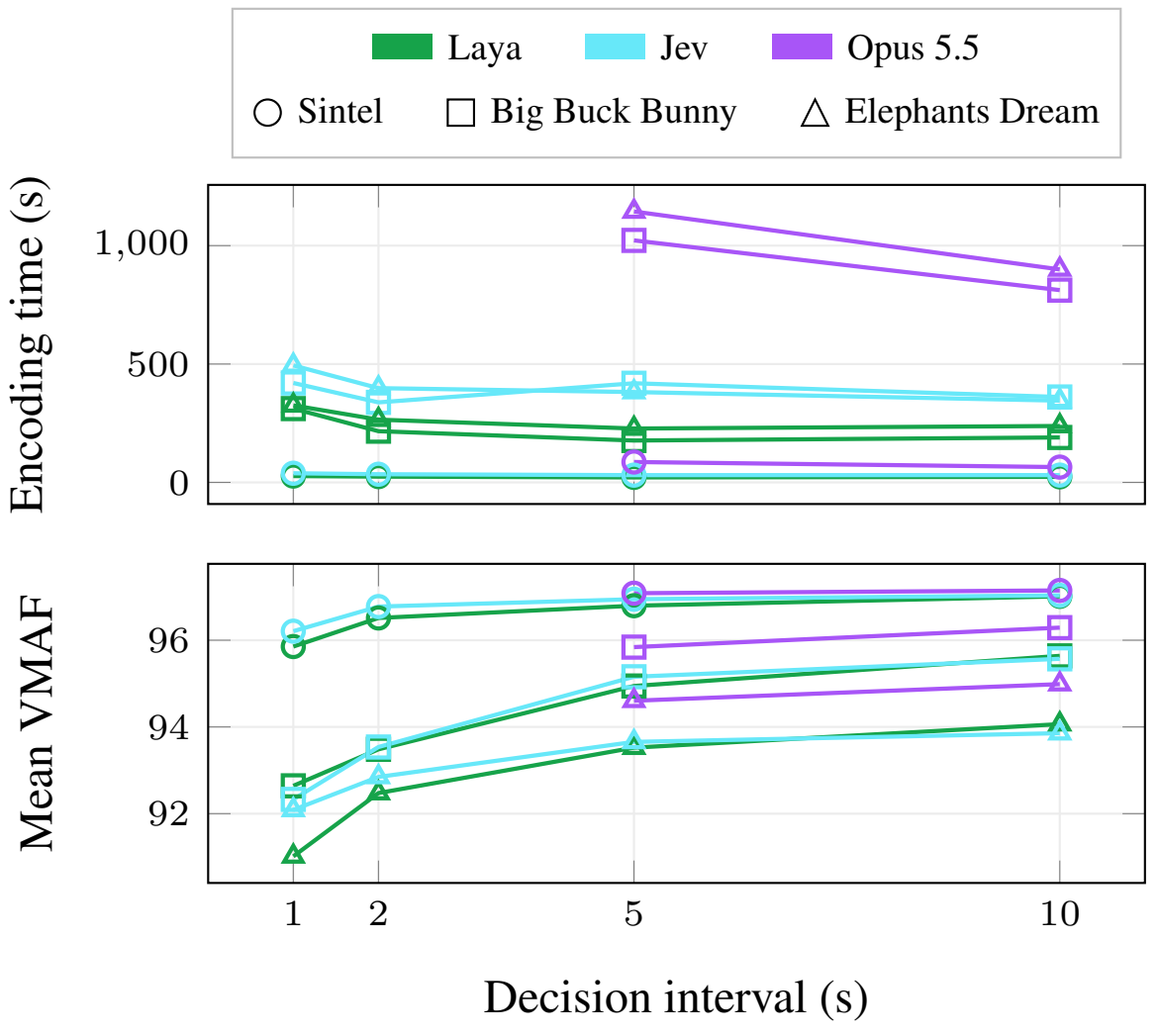
\begin{figure}[!t]
\centering
\makebox[\columnwidth][c]{%
\begin{tikzpicture}
\begin{groupplot}[
    group style={
        group size=1 by 2,
        vertical sep=0.4cm,
        x descriptions at=edge bottom
    },
    width=0.95\columnwidth,
    height=0.45\columnwidth,
    xlabel={Decision interval (s)},
    xtick={1,2,5,10},
    xmin=0,xmax=11,
    ylabel style={
        at={(axis description cs:0.05,0.5)},
        anchor=south,
    },
    grid=major,
    grid style={gray!15},
    tick label style={font=\scriptsize},
    label style={font=\small},
    scaled y ticks=false
]

\nextgroupplot[ylabel={Encoding time (s)}]
\PlotAblation{st}{laya}{wall_seconds}{layaColor}
\PlotAblation{bbb}{laya}{wall_seconds}{layaColor}
\PlotAblation{st}{jev}{wall_seconds}{jevColor}
\PlotAblation{bbb}{jev}{wall_seconds}{jevColor}
\PlotAblation{st}{opus}{wall_seconds}{opusColor}
\PlotAblation{bbb}{opus}{wall_seconds}{opusColor}
\PlotAblation{ed}{laya}{wall_seconds}{layaColor}
\PlotAblation{ed}{jev}{wall_seconds}{jevColor}
\PlotAblation{ed}{opus}{wall_seconds}{opusColor}

\nextgroupplot[ylabel={Mean VMAF}]
\PlotAblation{st}{laya}{vmaf_mean}{layaColor}
\PlotAblation{bbb}{laya}{vmaf_mean}{layaColor}
\PlotAblation{st}{jev}{vmaf_mean}{jevColor}
\PlotAblation{bbb}{jev}{vmaf_mean}{jevColor}
\PlotAblation{st}{opus}{vmaf_mean}{opusColor}
\PlotAblation{bbb}{opus}{vmaf_mean}{opusColor}
\PlotAblation{ed}{laya}{vmaf_mean}{layaColor}
\PlotAblation{ed}{jev}{vmaf_mean}{jevColor}
\PlotAblation{ed}{opus}{vmaf_mean}{opusColor}

% \nextgroupplot[ylabel={Compressed size (MB)}]
% \PlotAblation{st}{laya}{size_mb}{layaColor}
% \PlotAblation{bbb}{laya}{size_mb}{layaColor}
% \PlotAblation{st}{jev}{size_mb}{jevColor}
% \PlotAblation{bbb}{jev}{size_mb}{jevColor}
% \PlotAblation{st}{opus}{size_mb}{opusColor}
% \PlotAblation{bbb}{opus}{size_mb}{opusColor}
% \PlotAblation{st}{raw}{size_mb}{defaultColor}
% \PlotAblation{bbb}{raw}{size_mb}{defaultColor}

\end{groupplot}

% Single centered legend above all three plots.
% Two centered legends above all three plots.
\node[
    anchor=south,
    yshift=5pt,
    draw=black!25,
    fill=white,
    inner sep=4pt
] at (group c1r1.north) {% {%
    \scriptsize
    \begin{tabular}{@{}c@{}}
        % Encoder legend
        \begin{tabular}{@{}l@{\hspace{3pt}}l
                          @{\hspace{12pt}}l@{\hspace{3pt}}l
                          @{\hspace{12pt}}l@{\hspace{3pt}}l@{}}
            \tikz{\fill[layaColor] (0,-0.07) rectangle (0.4,0.1);}
            & Laya
            & \tikz{\fill[jevColor] (0,-0.07) rectangle (0.4,0.1);}
            & Jev
            & \tikz{\fill[opusColor] (0,-0.07) rectangle (0.4,0.1);}
            & Opus 5.5
        \end{tabular}
        \\[4pt]
        % Video legend
        \begin{tabular}{@{}l@{\hspace{3pt}}l
                          @{\hspace{12pt}}l@{\hspace{3pt}}l
                          @{\hspace{12pt}}l@{\hspace{3pt}}l@{}}
            \tikz[baseline=-2pt]{\draw (0,0) circle (2.5pt);}
            & Sintel
            & \tikz[baseline=0.5pt]{\draw (0,0) rectangle (5pt,5pt);}
            & Big Buck Bunny
            & \tikz[baseline=0.5pt]{\draw (0,0) -- (2.5pt,5pt) -- (5pt,0) -- cycle;}
            & Elephants Dream
        \end{tabular}
    \end{tabular}%
};
\end{tikzpicture}
}
\vspace{-0.8cm}
\caption{
Impact of decision interval, the time between encoding parameter updates, on encoding time and video quality for the three videos described in Table~\ref{tab:encoding-workloads}. Lower encoding time and higher VMAF are better.
}
\label{fig:video-ablation}
\end{figure}

%% file: eval_ffmpeg_results.tex
\subsubsection{Results}

Figure~\ref{fig:ffmpeg-tradeoff} compares encoding time and mean VMAF for the three videos. Laya achieves the lowest encoding time across all three. On Big Buck Bunny and Elephants Dream, it reduces encoding time by approximately 40--45\% relative to FFmpeg medium, with a reduction of less than two VMAF points. On Sintel, Laya achieves nearly the same quality as FFmpeg medium with a shorter encoding time.

We found that Opus 5.5 selects substantially more expensive encoder configurations than Laya. On Big Buck Bunny and Elephants Dream, Opus-Low runs 4.3× and 3.8× longer than Laya, for a mean VMAF gain of $<$1 point. Most of the extra time came from the heavier encoder settings Opus picks, not from model calls: 70--73\% of it is spent on encoding. For example, on Big Buck Bunny, Opus selects near-maximum encoding settings for 56 of 60 segments, while Laya selects faster motion estimation settings for approximately 45 segments. We also find little difference between Opus-Low and Opus-High: their VMAF scores differ by at most 0.04 points, while Opus-High takes 7--13\% longer.

%% file: eval_ffmpeg_ablation.tex
\subsubsection{Ablation Study}\label{subsec:ffmpeg-ablation}

Figure~\ref{fig:video-ablation} shows the effect of varying the interval between encoding-parameter decisions. Increasing the interval generally improves mean VMAF across the three videos by a few points, potentially too little to be perceptible to humans~\cite{ozer2017vmaf}, without a corresponding increase in encoding time. This suggests that more frequent decisions are not necessarily beneficial for video encoding, and that relatively infrequent adjustments can achieve favorable encoding-time--quality tradeoffs.

%% file: discussion.tex
\section{Discussion}

The benefit of additional reasoning varies across resource-management tasks. In microservice autoscaling, Opus with high reasoning effort substantially outperforms Opus with low reasoning effort, particularly on TeaStore. In contrast, additional reasoning provides little benefit in DRAM sizing and video encoding. In video encoding, despite explicit instructions to minimize encoding time subject to a quality target, Opus picks configurations that cost substantially more time than Laya's for slightly higher VMAF.

This motivates composing models with different capabilities for different decisions rather than relying on a single model: small, inexpensive decision models where additional reasoning showed no benefit, and more capable models where it did. In our autoscaling experiments, giving Jev application-wide state substantially improves its decisions, even though it still chooses replica counts independently for each service. Future work could explore how to divide decisions among models with different capabilities and coordinate their actions.

Although model outputs may be stochastic, restricting decisions to a finite set of choices allows us to treat the model as a nondeterministic component. This opens the possibility of using model checking to exhaustively explore possible actions and their interleavings, potentially providing formal guarantees for safety and liveness properties of a system. Verified safety constraints could then be enforced at runtime without prescribing how the model should decide.

%% file: related-work.tex
\section{Related Work}

\paragraph{AI-assisted policy design.}
A growing line of work uses LLMs to search for algorithms and heuristics: FunSearch~\cite{funsearch} uses LLMs to discover mathematical constructions and bin-packing heuristics through iterative generation and evaluation. Vulcan~\cite{vulcan} generates heuristics tailored to specific systems. The ADRS framework~\cite{barbarians} applies the generate--evaluate--refine loop across scheduling, load balancing, and LLM-based SQL processing. Finally, AI-greement~\cite{agreement} uses agents to discover and formally verify a consensus protocol. In each case, the model's output is an artifact---a program, a heuristic, a protocol---that is evaluated and subsequently executes without the model. In contrast, we use pretrained models to make resource-management decisions directly at runtime, rather than develop an algorithm or policy for making them.

\paragraph{Task-specific resource management.}
Most prior work on microservice resource management focuses on \emph{vertical} scaling, adjusting CPU and memory allocations per service~\cite{sinan-delimitrou,firm-osdi,autothrottle,percs}. Horizontal autoscaling instead adjusts replica counts, using threshold-based controllers~\cite{k8s-hpa, gke-metric-autoscaling,gcp-autoscaler,aws-autoscaling}, bandit-based approaches~\cite{erlang}, or analytical what-if models~\cite{agar}. In tiered memory, systems such as TPP~\cite{tpp2023}, HeMem~\cite{hemem2021} and ARMS~\cite{arms} use policies to place frequently accessed pages in DRAM, while TierSizer~\cite{tiersizer} uses analytical models to determine DRAM allocations at runtime. Pond~\cite{cxlpond} predicts DRAM allocations at VM startup and falls back to fully local DRAM if performance degrades, rather than continuously adjusting allocations at runtime. Vulcan~\cite{tang2025leave} dynamically partitions a fixed DRAM budget among co-running workloads to improve fairness and efficiency, rather than sizing DRAM to meet individual performance targets. Tuna~\cite{tuna} dynamically sizes DRAM using offline-profiled microbenchmarks, but relies on predefined performance models that may generalize poorly to unseen workloads.

Video encoding involves tradeoffs between encoding time, compression efficiency, and visual quality. Prior work characterizes these tradeoffs across encoder configurations~\cite{msu2008x264} and selects among predefined encoding presets, each specifying a collection of encoder parameter settings~\cite{amirpour2023preset}. Content-adaptive approaches construct bitrate ladders by evaluating candidate encodings~\cite{netflixDynamicOptimizer} or predicting bitrate--resolution combinations from video features~\cite{ladder-prediction}. We select individual encoder parameters rather than presets, and adjust them per segment from observed encoding time and quality.

\paragraph{Learned models as the policy.}
Reinforcement learning has been used to train runtime policies for cluster scheduling~\cite{decima}, adaptive-bitrate streaming~\cite{pensieve}, and congestion control~\cite{aurora}. These systems learn decision policies from task-specific experience, often collected through simulation, using rewards that encode the desired behavior. Although the resulting policies can generalize to unseen workloads and environments, they still require training for their respective tasks. \name models receive no such training; they act on the current state, objective, and available actions alone.

%% file: conclusion.tex
\section{Conclusion}

Resource management has traditionally relied on task-specific policies that map system measurements to actions. We investigate whether explicitly developing such policies is necessary. Across microservice autoscaling, DRAM sizing, and video encoding, we find that pretrained models can make competitive decisions directly from observed system state, objectives, and available choices, without task-specific training or policy development. Small decision models achieve results comparable to a frontier LLM and policy-based systems in several cases, though not always within performance targets. Our results challenge the assumption that effective resource management must begin with developing a task-specific policy.

\begin{center}
\emph{This work does not raise any ethical questions.}
\end{center}

%% file: main.bbl
\begin{thebibliography}{58}
\providecommand{\natexlab}[1]{#1}
\providecommand{\url}[1]{\texttt{#1}}
\expandafter\ifx\csname urlstyle\endcsname\relax
  \providecommand{\doi}[1]{doi: #1}\else
  \providecommand{\doi}{doi: \begingroup \urlstyle{rm}\Url}\fi

\bibitem[Achermann \& Panwar(2020)Achermann and Panwar]{reto2020btree}
Achermann, R. and Panwar, A.
\newblock Mitosis workload btree.
\newblock 2020.
\newblock URL \url{{https://github.com/mitosis-project/mitosis-workload-btree}}.

\bibitem[Almeida(2026)]{jev}
Almeida, D.
\newblock Introducing {System One Models} \& {Jev}.
\newblock TypeSafe AI Blog, September 2026.
\newblock URL \url{https://typesafe.ai/blog/introducing-system-one-models-and-jev}.
\newblock Accessed: 2026-10-24.

\bibitem[{Amazon Web Services}(2024)]{aws-autoscaling}
{Amazon Web Services}.
\newblock Target tracking scaling policies for {Amazon} {EC2} auto scaling.
\newblock \url{https://docs.aws.amazon.com/autoscaling/ec2/userguide/as-scaling-target-tracking.html}, 2024.
\newblock Accessed: 2025-06-20.

\bibitem[Amirpour et~al.(2023)Amirpour, Menon, Afzal, Prodan, and Timmerer]{amirpour2023preset}
Amirpour, H., Menon, V.~V., Afzal, S., Prodan, R., and Timmerer, C.
\newblock Optimizing video streaming for sustainability and quality: The role of preset selection in per-title encoding.
\newblock In \emph{IEEE International Conference on Multimedia and Expo (ICME)}, pp.\  1679--1684, 2023.
\newblock \doi{10.1109/ICME55011.2023.00289}.

\bibitem[{Anthropic}(2026)]{opus-5-5}
{Anthropic}.
\newblock Introducing claude opus 5.5.
\newblock \url{https://www.anthropic.com/claude-opus-5-5}, 2026.
\newblock Accessed: 2026-10-30.

\bibitem[Beamer et~al.(2015)Beamer, Asanovi{\'c}, and Patterson]{beamer2015gap}
Beamer, S., Asanovi{\'c}, K., and Patterson, D.
\newblock The gap benchmark suite.
\newblock \emph{arXiv preprint arXiv:1508.03619}, 2015.

\bibitem[{Blender Foundation}(2006)]{elephants-dream}
{Blender Foundation}.
\newblock {Elephants Dream}.
\newblock \url{https://orange.blender.org/}, 2006.

\bibitem[{Blender Foundation}(2008)]{big-buck-bunny}
{Blender Foundation}.
\newblock {Big Buck Bunny}.
\newblock \url{https://peach.blender.org/}, 2008.

\bibitem[{Blender Foundation}(2010)]{sintel}
{Blender Foundation}.
\newblock {Sintel}.
\newblock \url{https://durian.blender.org/}, 2010.

\bibitem[Burgelin et~al.(2026)Burgelin, Guerraoui, Hong, Hu, Kermarrec, Pires, Veski, and de~Vos]{agreement}
Burgelin, C., Guerraoui, R., Hong, Z., Hu, Y., Kermarrec, A.-M., Pires, R., Veski, R.~M., and de~Vos, M.
\newblock The end of consensus research?, 2026.
\newblock URL \url{https://arxiv.org/abs/2610.30114}.

\bibitem[Chen et~al.(2024)Chen, Huang, Yang, Liu, Li, Nikolopoulos, Ryu, Baek, Shin, and Li]{tuna}
Chen, S., Huang, J., Yang, S., Liu, J., Li, H., Nikolopoulos, D., Ryu, J., Baek, J., Shin, K., and Li, D.
\newblock Tuning fast memory size based on modeling of page migration for tiered memory.
\newblock \emph{arXiv preprint arXiv:2410.00328}, 2024.

\bibitem[Cheng et~al.(2025)Cheng, Liu, Pan, Li, Wang, Krentsel, Xia, Cemri, Park, Yang, Chen, Agrawal, Desai, Xing, Sen, Zaharia, and Stoica]{barbarians}
Cheng, A., Liu, S., Pan, M., Li, Z., Wang, B., Krentsel, A., Xia, T., Cemri, M., Park, J., Yang, S., Chen, J., Agrawal, L., Desai, A., Xing, J., Sen, K., Zaharia, M., and Stoica, I.
\newblock Barbarians at the gate: How ai is upending systems research, 2025.
\newblock URL \url{https://arxiv.org/abs/2510.06189}.

\bibitem[{Convai Innovations}()]{laya-huggingface}
{Convai Innovations}.
\newblock Laya.
\newblock \url{https://huggingface.co/convaiinnovations/laya}.

\bibitem[{Convai Innovations}(2026)]{laya}
{Convai Innovations}.
\newblock {Laya}.
\newblock Hugging Face model card, 2026.
\newblock URL \url{https://huggingface.co/convaiinnovations/laya}.
\newblock Accessed: 2026-10-24.

\bibitem[{Datadog}(2026)]{datadog-containers-serverless-2025}
{Datadog}.
\newblock State of containers and serverless.
\newblock \url{https://www.datadoghq.com/state-of-containers-and-serverless/}, 2026.
\newblock Accessed: 2026-09-03.

\bibitem[Deshpande et~al.(2026)Deshpande, Yadwadkar, and Arun]{agar}
Deshpande, S., Yadwadkar, N.~J., and Arun, V.
\newblock Agar: What-if analysis for microservice applications, 2026.
\newblock URL \url{https://arxiv.org/abs/2610.36636}.

\bibitem[Duplyakin et~al.(2019)Duplyakin, Ricci, Maricq, Wong, Duerig, Eide, Stoller, Hibler, Johnson, Webb, Akella, Wang, Ricart, Landweber, Elliott, Zink, Cecchet, Kar, and Mishra]{cloudlab}
Duplyakin, D., Ricci, R., Maricq, A., Wong, G., Duerig, J., Eide, E., Stoller, L., Hibler, M., Johnson, D., Webb, K., Akella, A., Wang, K., Ricart, G., Landweber, L., Elliott, C., Zink, M., Cecchet, E., Kar, S., and Mishra, P.
\newblock The design and operation of {CloudLab}.
\newblock In \emph{Proceedings of the {USENIX} Annual Technical Conference (ATC)}, pp.\  1--14, July 2019.
\newblock URL \url{https://www.flux.utah.edu/paper/duplyakin-atc19}.

\bibitem[Dwivedula et~al.(2025)Dwivedula, Saxena, Yadalam, Campbell, Kim, and Akella]{vulcan}
Dwivedula, R., Saxena, D., Yadalam, S., Campbell, E.~H., Kim, D., and Akella, A.
\newblock Vulcan: Instance-specialized, verifiable systems heuristics through llm-driven search.
\newblock \emph{arXiv preprint arXiv:2512.25065}, 2025.

\bibitem[Fan et~al.(2008)Fan, Chang, Hsieh, Wang, and Lin]{fan2008liblinear}
Fan, R.-E., Chang, K.-W., Hsieh, C.-J., Wang, X.-R., and Lin, C.-J.
\newblock Liblinear: A library for large linear classification.
\newblock \emph{The Journal of machine Learning research}, 2008.

\bibitem[Gan et~al.(2019)Gan, Zhang, Cheng, Shetty, Rathi, Katarki, Bruno, Hu, Ritchken, Jackson, Hu, Pancholi, He, Clancy, Colen, Wen, Leung, Wang, Zaruvinsky, Espinosa, Lin, Liu, Padilla, and Delimitrou]{deathstarbench}
Gan, Y., Zhang, Y., Cheng, D., Shetty, A., Rathi, P., Katarki, N., Bruno, A., Hu, J., Ritchken, B., Jackson, B., Hu, K., Pancholi, M., He, Y., Clancy, B., Colen, C., Wen, F., Leung, C., Wang, S., Zaruvinsky, L., Espinosa, M., Lin, R., Liu, Z., Padilla, J., and Delimitrou, C.
\newblock An open-source benchmark suite for microservices and their hardware-software implications for cloud \& edge systems.
\newblock In \emph{Proceedings of the Twenty-Fourth International Conference on Architectural Support for Programming Languages and Operating Systems}, ASPLOS '19, pp.\  3–18, New York, NY, USA, 2019. Association for Computing Machinery.
\newblock ISBN 9781450362405.
\newblock \doi{10.1145/3297858.3304013}.
\newblock URL \url{https://doi.org/10.1145/3297858.3304013}.

\bibitem[{Google Cloud}(2024)]{gcp-autoscaler}
{Google Cloud}.
\newblock Autoscaling groups of instances.
\newblock \url{https://cloud.google.com/compute/docs/autoscaler}, 2024.
\newblock Accessed: 2025-06-20.

\bibitem[{Google Cloud}(2026)]{gke-metric-autoscaling}
{Google Cloud}.
\newblock About autoscaling workloads based on metrics.
\newblock \url{https://docs.cloud.google.com/kubernetes-engine/docs/concepts/custom-and-external-metrics}, 2026.
\newblock Google Kubernetes Engine documentation, last updated 2026-03-26, accessed 2026-03-30.

\bibitem[GoogleCloudPlatform(2020)]{online-boutique-github}
GoogleCloudPlatform.
\newblock Online {Boutique}: A cloud-native microservices demo application.
\newblock \url{https://github.com/GoogleCloudPlatform/microservices-demo}, 2020.

\bibitem[{Intel Corporation.}(2018)]{pebs2018}
{Intel Corporation.}
\newblock Intel 64 and ia-32 architectures software developer manuals.
\newblock 2018.
\newblock URL \url{{https://software.intel.com/articles/intel-sdm}}.

\bibitem[Jay et~al.(2019)Jay, Rotman, Godfrey, Schapira, and Tamar]{aurora}
Jay, N., Rotman, N.~H., Godfrey, P.~B., Schapira, M., and Tamar, A.
\newblock A deep reinforcement learning perspective on internet congestion control.
\newblock In \emph{Proceedings of the 36th International Conference on Machine Learning (ICML)}, volume~97 of \emph{Proceedings of Machine Learning Research}, pp.\  3050--3059. PMLR, 2019.

\bibitem[{Jev AI}()]{jev-models}
{Jev AI}.
\newblock Jev models.
\newblock \url{https://jev-ai.org/docs/models/}.

\bibitem[Katsenou et~al.(2021)Katsenou, Sole, and Bull]{ladder-prediction}
Katsenou, A.~V., Sole, J., and Bull, D.~R.
\newblock Efficient bitrate ladder construction for content-optimized adaptive video streaming.
\newblock \emph{IEEE Open Journal of Signal Processing}, 2:\penalty0 496--511, 2021.
\newblock \doi{10.1109/OJSP.2021.3086691}.

\bibitem[{Knative Project}(2024)]{knative-kpa}
{Knative Project}.
\newblock Autoscaling concepts.
\newblock \url{https://knative.dev/docs/serving/autoscaling/autoscaling-concepts/}, 2024.
\newblock Accessed: 2025-06-20.

\bibitem[{Kubernetes}(2024)]{k8s-hpa}
{Kubernetes}.
\newblock Horizontal pod autoscaler.
\newblock \url{https://kubernetes.io/docs/tasks/run-application/horizontal-pod-autoscale/}, 2024.
\newblock Accessed: 2025-06-20.

\bibitem[Lee et~al.(2023)Lee, Monga, Min, and Eom]{memtis2023}
Lee, T., Monga, S.~K., Min, C., and Eom, Y.~I.
\newblock Memtis: Efficient memory tiering with dynamic page classification and page size determination.
\newblock In \emph{Proceedings of the 29th Symposium on Operating Systems Principles}, 2023.

\bibitem[Lepers \& Zwaenepoel(2023)Lepers and Zwaenepoel]{johnnyCache2023}
Lepers, B. and Zwaenepoel, W.
\newblock Johnny cache: the end of $\{$DRAM$\}$ cache conflicts (in tiered main memory systems).
\newblock In \emph{17th USENIX Symposium on Operating Systems Design and Implementation}, 2023.

\bibitem[Li et~al.(2023)Li, Berger, Hsu, Ernst, Zardoshti, Novakovic, Shah, Rajadnya, Lee, Agarwal, Hill, Fontoura, and Bianchini]{cxlpond}
Li, H., Berger, D.~S., Hsu, L., Ernst, D., Zardoshti, P., Novakovic, S., Shah, M., Rajadnya, S., Lee, S., Agarwal, I., Hill, M.~D., Fontoura, M., and Bianchini, R.
\newblock Pond: Cxl-based memory pooling systems for cloud platforms.
\newblock In \emph{Proceedings of the 28th ACM International Conference on Architectural Support for Programming Languages and Operating Systems, Volume 2}, ASPLOS 2023, pp.\  574–587, New York, NY, USA, 2023. Association for Computing Machinery.
\newblock ISBN 9781450399166.
\newblock \doi{10.1145/3575693.3578835}.
\newblock URL \url{https://doi.org/10.1145/3575693.3578835}.

\bibitem[Li et~al.(2016)Li, Aaron, Katsavounidis, Moorthy, and Manohara]{vmaf}
Li, Z., Aaron, A., Katsavounidis, I., Moorthy, A., and Manohara, M.
\newblock Toward a practical perceptual video quality metric, June 2016.
\newblock URL \url{https://netflixtechblog.com/toward-a-practical-perceptual-video-quality-metric-653f208b9652}.

\bibitem[{Locust team}(ongoing)]{locust}
{Locust team}.
\newblock {Locust}: A modern load testing framework, ongoing.
\newblock \url{https://locust.io/}.

\bibitem[Luo et~al.(2021)Luo, Xu, Lu, Ye, Xu, Zhang, Ding, He, and Xu]{alibaba-2021}
Luo, S., Xu, H., Lu, C., Ye, K., Xu, G., Zhang, L., Ding, Y., He, J., and Xu, C.
\newblock Characterizing microservice dependency and performance: Alibaba trace analysis.
\newblock In \emph{Proceedings of the ACM Symposium on Cloud Computing}, pp.\  412--426, 2021.

\bibitem[Mao et~al.(2017)Mao, Netravali, and Alizadeh]{pensieve}
Mao, H., Netravali, R., and Alizadeh, M.
\newblock Neural adaptive video streaming with {Pensieve}.
\newblock In \emph{Proceedings of the Conference of the ACM Special Interest Group on Data Communication (SIGCOMM)}, pp.\  197--210, 2017.
\newblock \doi{10.1145/3098822.3098843}.

\bibitem[Mao et~al.(2019)Mao, Schwarzkopf, Venkatakrishnan, Meng, and Alizadeh]{decima}
Mao, H., Schwarzkopf, M., Venkatakrishnan, S.~B., Meng, Z., and Alizadeh, M.
\newblock Learning scheduling algorithms for data processing clusters.
\newblock In \emph{Proceedings of the ACM Special Interest Group on Data Communication (SIGCOMM '19)}, pp.\  270--288. ACM, 2019.
\newblock \doi{10.1145/3341302.3342080}.

\bibitem[Maruf et~al.(2023)Maruf, Wang, Dhanotia, Weiner, Agarwal, Bhattacharya, Petersen, Chowdhury, Kanaujia, and Chauhan]{tpp2023}
Maruf, H.~A., Wang, H., Dhanotia, A., Weiner, J., Agarwal, N., Bhattacharya, P., Petersen, C., Chowdhury, M., Kanaujia, S., and Chauhan, P.
\newblock Tpp: Transparent page placement for cxl-enabled tiered-memory.
\newblock In \emph{Proceedings of the 28th ACM International Conference on Architectural Support for Programming Languages and Operating Systems}, 2023.

\bibitem[{MSU Graphics \& Media Lab}(2008)]{msu2008x264}
{MSU Graphics \& Media Lab}.
\newblock Options analysis of mpeg-4 avc/h.264 codec x264.
\newblock \url{https://www.compression.ru/video/codec_comparison/x264_options_analysis_08_en.html}, 2008.

\bibitem[{Netflix TechBlog}(2018)]{netflixDynamicOptimizer}
{Netflix TechBlog}.
\newblock Dynamic optimizer: A perceptual video encoding optimization framework.
\newblock \url{https://netflixtechblog.com/dynamic-optimizer-a-perceptual-video-encoding-optimization-framework-e19f1e3a277f}, 2018.

\bibitem[Ozer(2017)]{ozer2017vmaf}
Ozer, J.
\newblock Finding the just noticeable difference with netflix vmaf.
\newblock Streaming Learning Center, 2017.
\newblock URL \url{https://streaminglearningcenter.com/blogs/finding-the-just-noticeable-difference-with-netflix-vmaf.html}.

\bibitem[Plimpton et~al.(2006)Plimpton, Brightwell, Vaughan, Underwood, and Davis]{plimpton2006simple}
Plimpton, S.~J., Brightwell, R., Vaughan, C., Underwood, K., and Davis, M.
\newblock A simple synchronous distributed-memory algorithm for the hpcc randomaccess benchmark.
\newblock In \emph{2006 IEEE International Conference on Cluster Computing}, 2006.

\bibitem[Qiu et~al.(2020)Qiu, Banerjee, Jha, Kalbarczyk, and Iyer]{firm-osdi}
Qiu, H., Banerjee, S.~S., Jha, S., Kalbarczyk, Z.~T., and Iyer, R.~K.
\newblock {FIRM}: An intelligent fine-grained resource management framework for {SLO-Oriented} microservices.
\newblock In \emph{14th USENIX Symposium on Operating Systems Design and Implementation (OSDI 20)}, pp.\  805--825. USENIX Association, November 2020.
\newblock ISBN 978-1-939133-19-9.
\newblock URL \url{https://www.usenix.org/conference/osdi20/presentation/qiu}.

\bibitem[Raybuck et~al.(2021)Raybuck, Stamler, Zhang, Erez, and Peter]{hemem2021}
Raybuck, A., Stamler, T., Zhang, W., Erez, M., and Peter, S.
\newblock Hemem: Scalable tiered memory management for big data applications and real nvm.
\newblock In \emph{Proceedings of the ACM SIGOPS 28th Symposium on Operating Systems Principles}, 2021.

\bibitem[Romera-Paredes et~al.(2024)Romera-Paredes, Barekatain, Novikov, Balog, Kumar, Dupont, Ruiz, Ellenberg, Wang, Fawzi, Kohli, and Fawzi]{funsearch}
Romera-Paredes, B., Barekatain, M., Novikov, A., Balog, M., Kumar, M.~P., Dupont, E., Ruiz, F. J.~R., Ellenberg, J.~S., Wang, P., Fawzi, O., Kohli, P., and Fawzi, A.
\newblock Mathematical discoveries from program search with large language models.
\newblock \emph{Nature}, 625\penalty0 (7995):\penalty0 468--475, January 2024.
\newblock \doi{10.1038/s41586-023-06924-6}.
\newblock URL \url{https://www.nature.com/articles/s41586-023-06924-6}.

\bibitem[Sachidananda \& Sivaraman(2024)Sachidananda and Sivaraman]{erlang}
Sachidananda, V. and Sivaraman, A.
\newblock Erlang: Application-aware autoscaling for cloud microservices.
\newblock In \emph{Proceedings of the Nineteenth European Conference on Computer Systems}, EuroSys '24, pp.\  888–923, New York, NY, USA, 2024. Association for Computing Machinery.
\newblock ISBN 9798400704376.
\newblock \doi{10.1145/3627703.3650084}.
\newblock URL \url{https://doi.org/10.1145/3627703.3650084}.

\bibitem[Saxena et~al.(2026)Saxena, Vipat, Lin, Wang, Dillig, Shakkottai, and Akella]{percs}
Saxena, D., Vipat, G., Lin, J., Wang, J., Dillig, I., Shakkottai, S., and Akella, A.
\newblock Towards performance robustness for microservices.
\newblock In \emph{23rd USENIX Symposium on Networked Systems Design and Implementation (NSDI 26)}, pp.\  93--112, Renton, WA, May 2026. USENIX Association.
\newblock ISBN 978-1-939133-54-0.
\newblock URL \url{https://www.usenix.org/conference/nsdi26/presentation/saxena}.

\bibitem[Sullivan \& Wiegand(1998)Sullivan and Wiegand]{param-optimization-video-compression}
Sullivan, G.~J. and Wiegand, T.
\newblock Rate-distortion optimization for video compression.
\newblock \emph{IEEE Signal Processing Magazine}, 15\penalty0 (6):\penalty0 74--90, 1998.
\newblock \doi{10.1109/79.733497}.

\bibitem[Tang et~al.(2025)Tang, Wang, Wang, and Wu]{tang2025leave}
Tang, W., Wang, Y., Wang, Y., and Wu, J.
\newblock Leave no one behind: Fair and efficient tiered memory management for multi-applications.
\newblock In \emph{Proceedings of the 54th International Conference on Parallel Processing}, pp.\  699--709, 2025.

\bibitem[Tramm et~al.(2014)Tramm, Siegel, Islam, and Schulz]{tramm2014xsbench}
Tramm, J.~R., Siegel, A.~R., Islam, T., and Schulz, M.
\newblock {XSBench}: The development and verification of a performance abstraction for monte carlo reactor analysis.
\newblock In \emph{PHYSOR 2014: The Role of Reactor Physics toward a Sustainable Future}, 2014.
\newblock URL \url{https://www.mcs.anl.gov/papers/P5064-0114.pdf}.

\bibitem[Tu et~al.(2013)Tu, Zheng, Kohler, Liskov, and Madden]{tu2013speedy}
Tu, S., Zheng, W., Kohler, E., Liskov, B., and Madden, S.
\newblock Speedy transactions in multicore in-memory databases.
\newblock In \emph{Proceedings of the Twenty-Fourth ACM Symposium on Operating Systems Principles}, 2013.

\bibitem[von Kistowski et~al.(2018)von Kistowski, Eismann, Schmitt, Bauer, Grohmann, and Kounev]{teastore}
von Kistowski, J., Eismann, S., Schmitt, N., Bauer, A., Grohmann, J., and Kounev, S.
\newblock {TeaStore: A Micro-Service Reference Application for Benchmarking, Modeling and Resource Management Research}.
\newblock In \emph{Proceedings of the 26th IEEE International Symposium on the Modelling, Analysis, and Simulation of Computer and Telecommunication Systems}, MASCOTS '18, September 2018.

\bibitem[Wang et~al.(2024)Wang, Li, Liang, Wu, and Yan]{autothrottle}
Wang, Z., Li, P., Liang, C.-J.~M., Wu, F., and Yan, F.~Y.
\newblock Autothrottle: A practical {Bi-Level} approach to resource management for {SLO-Targeted} microservices.
\newblock In \emph{21st USENIX Symposium on Networked Systems Design and Implementation (NSDI 24)}, pp.\  149--165, Santa Clara, CA, April 2024. USENIX Association.
\newblock ISBN 978-1-939133-39-7.
\newblock URL \url{https://www.usenix.org/conference/nsdi24/presentation/wang-zibo}.

\bibitem[{Xiph.Org Foundation}()]{xiph-media}
{Xiph.Org Foundation}.
\newblock Xiph.org video test media.
\newblock \url{https://media.xiph.org/}.
\newblock Accessed: 2026-10-29.

\bibitem[Yadalam et~al.(2025)Yadalam, Kanellis, Swift, and Venkataraman]{arms}
Yadalam, S., Kanellis, K., Swift, M., and Venkataraman, S.
\newblock Arms: Adaptive and robust memory tiering system.
\newblock \emph{arXiv preprint arXiv:2508.04417}, 2025.

\bibitem[Yadalam et~al.(2026)Yadalam, Deshpande, Saxena, Arun, Yadwadkar, and Akella]{tiersizer}
Yadalam, S., Deshpande, S., Saxena, D., Arun, V., Yadwadkar, N.~J., and Akella, A.
\newblock {TierSizer}: Counterfactual reasoning for {DRAM} sizing in tiered memory systems.
\newblock In \emph{4th Workshop on Disruptive Memory Systems ({DIMES})}, 2026.

\bibitem[Yang et~al.(2017)Yang, Hoseinzadeh, Andrews, Mayers, Evans, Bolt, Bhimani, Mi, and Swanson]{autotiering2017}
Yang, Z., Hoseinzadeh, M., Andrews, A., Mayers, C., Evans, D.~T., Bolt, R.~T., Bhimani, J., Mi, N., and Swanson, S.
\newblock Autotiering: Automatic data placement manager in multi-tier all-flash datacenter.
\newblock In \emph{2017 IEEE 36th International Performance Computing and Communications Conference}, 2017.

\bibitem[Zhang et~al.(2021)Zhang, Hua, Zhou, Suh, and Delimitrou]{sinan-delimitrou}
Zhang, Y., Hua, W., Zhou, Z., Suh, G.~E., and Delimitrou, C.
\newblock Sinan: Ml-based and qos-aware resource management for cloud microservices.
\newblock In \emph{Proceedings of the 26th ACM International Conference on Architectural Support for Programming Languages and Operating Systems}, ASPLOS '21, pp.\  167–181, New York, NY, USA, 2021. Association for Computing Machinery.
\newblock ISBN 9781450383172.
\newblock \doi{10.1145/3445814.3446693}.
\newblock URL \url{https://doi.org/10.1145/3445814.3446693}.

\end{thebibliography}
